\documentclass[a4paper,10pt,twocolumn,nofootinbib]{revtex4-1}
\usepackage{amsmath}
\usepackage{amsthm}
\usepackage{amssymb}
\usepackage[utf8]{inputenc}
\usepackage{graphicx}
\usepackage{tikz}
\usepackage{hyperref}
\hypersetup{pdftex,colorlinks=true,allcolors=blue}
\usepackage{hypcap}
\usepackage{pgfplots}
\usepackage{enumitem}
\usepackage[normalem]{ulem}
\usepackage{moreenum}

\usetikzlibrary{decorations.markings}
\tikzset{->-/.style={decoration={
  markings,
  mark=at position #1 with {\arrow{>}}},postaction={decorate}}}
\tikzset{-<-/.style={decoration={
  markings,
  mark=at position #1 with {\arrow{<}}},postaction={decorate}}}

\usepackage{textgreek}
\usepackage[outline]{contour}
\def\defn#1{{\bf #1}}

\def\Real{{\mathbb R}}

\def\norm#1{\left|#1\right|}
\def\innerprod(#1,#2){{\left<#1\,,\,#2\right>}}
\def\Set#1{{\left\{#1\right\}}}
\def\qquadtext#1{\qquad\textup{#1}\qquad}
\def\qquadand{\qquadtext{and}}
\def\quadtext#1{\quad\textup{#1}\quad}
\def\quadand{\quadtext{and}}

\usepackage{fancyhdr}
\fancypagestyle{firststyle}
{
   \fancyhf{}
   
\fancyfoot[C]{1}
\fancyfoot[R]{\footnotesize\bf File: \jobname.tex}
}
\newenvironment{bulletlist}{\begin{list}{$\bullet$}
{
\setlength{\itemindent}{0 em}
\setlength{\itemsep}{1pt}
\setlength{\labelsep}{0.5em}
\setlength{\labelwidth}{15em}
\setlength{\leftmargin}{1.5em}
\setlength{\parsep}{1pt}
\setlength{\parskip}{0em}
\setlength{\partopsep}{0pt}
\setlength{\topsep}{5pt}}}
{\end{list}}

\def\cRed#1{#1}

\def\Mman{{\mathcal{M}}}
\def\Je{{J}}
\def\Fem{{F}}
\def\Aem{{A}}

\def\axi{{\zeta}}
\def\axiTop{{\zeta_{\textup{top}}}}
\def\axiSource{{Z}_\textup{top}}
\def\axiTrueSource{{\xi}_\textup{top}}
\def\axiSE{{Z}_\textup{SE}}
\def\axiSB{{Z}_\textup{SB}}
\def\StanAxiSource{{Z_{\textup{std}}}}
\def\StanAxi{{\phi_{\textup{std}}}}
\def\StanAxizeta{{\zeta_{\textup{std}}}}
\def\FutLC{{\cal J}_+}

\def\tHat{{\hat{t}}}
\def\xhat{{\hat{x}}}
\def\yhat{{\hat{y}}}
\def\zhat{{\hat{z}}}
\def\Fhat{{\hat{\Fem}}}
\def\Jhat{{\hat{\Je}}}

\def\Mmanhat{\hat{\mathcal{M}}}

\def\axihat{\hat{\zeta}}
\def\FutLChat{\hat{\cal J}_+}

\DeclareMathOperator{\starhat}{{\hat\star}}

\def\MmanSup{{\Mman_{\textup{sup}}}}

\def\funr{{f_1}}
\def\funz{{f_2}}
\def\funA{{f_3}}

\def\VAct#1{\left<#1\right>}

\newcommand{\pVec}[1]{\boldsymbol{#1}}

\def\pE{E} % a scalar electric field
\def\pB{B} % a scalar magnetic field
\def\VE{\pVec{\pE}} % the EM electric field
\def\VB{\pVec{\pB}} % the EM magnetic field
\def\VD{\pVec{D}} % the EM electric excitation field (displacement field)
\def\VH{\pVec{H}} % the EM magnetic excitation field
\def\pini{{\boldsymbol p}_0}
\def\pfin{{\boldsymbol p}_1}
\def\ghat{{\hat g}}
\def\interval{{\cal I}}
\def\intvHat{\hat{\cal I}}
\def\axihat{{\hat\zeta}}
\def\axiSourcehat{{\hat\axiSource}}
\def\axiTrueSource{{\xi}_\textup{top}}

\def\manU{{\mathcal{U}}}
\def\manN{{\mathcal{N}}}

\def\SigmaPast{{\Sigma_-}}
\def\SigmaFut{{\Sigma_+}}
\def\QFut{{Q_+}}
\def\Qax{{Q_{\textup{ax}}}}
\def\QF{{Q_{\textup{EM}}}}

\def\SigmaFutHat{{\hat{\Sigma}_+}}
\def\QFutHat{{\hat{Q}_+}}
\def\manUHat{\hat{\cal U}}

\def\axiName{\cRed{field}}
\def\axiNameSource{\cRed{flux}}

\def\colLiCone{blue}             % light cone surface colour (a label)
\def\colInCone{green}            % light cone interior (a label)
\def\colUsurf{purple}            % U surface colour (a label)
\def\colPsing{red}               % singularity colour (a label)
\begin{document}
%\jgpapersize
\title{Temporary Singularities and Axions: \\
 an analytic solution that challenges charge conservation}

\author{Jonathan Gratus$^{1,2}$}
\homepage[]{https://orcid.org/0000-0003-1597-6084}
\email[\hphantom{.}~]{j.gratus@lancaster.ac.uk}
\author{Paul Kinsler$^{1,2,3}$}
\homepage[]{https://orcid.org/0000-0001-5744-8146}
\email[\hphantom{.}~]{Dr.Paul.Kinsler@physics.org}
\author{Martin W. McCall$^3$}
\homepage[]{https://orcid.org/0000-0003-0643-7169}
\email[\hphantom{.}~]{m.mccall@imperial.ac.uk}

\affiliation{$^1$
  Department of Physics,
  Lancaster University,
  Lancaster LA1 4YB,
  United Kingdom,
}
\affiliation{$^2$
The Cockcroft Institute,
Sci-Tech Daresbury,
Daresbury WA4 4AD,
% Warrington WA4 4AD
United Kingdom,
}
\affiliation{$^3$
  %Blackett Laboratory, Imperial College,
  Department of Physics,
  Imperial College London,
  Prince Consort Road,
  London SW7 2AZ,
  United Kingdom.
}

\begin{abstract}

We construct 
 an analytic solution 
 for electromagnetic fields interacting with an axion {\axiName}
 that
 violates global charge conservation, 
 by building on the possibilities demonstrated in 
 [\href{https://doi.org/10.1007/s10701-019-00251-5}{Foundations of Physics 49, 330 (2019)}]. 
Despite providing a specific example where 
 ``physics breaks down'' at a singularity,
 it nevertheless demonstrates that  
 the physical laws on the surrounding spacetime
 still impose constraints
 on what is allowed to happen.
The construction
 is valid for
 a spacetime containing a temporary singularity
 and a Maxwellian electrodynamics
 containing a {proposed} ``topological'' axion field.
Further,
 the concepts of transformation optics 
 can be applied 
 to show that our
 specific mathematical solution has a much wider applicability.

\end{abstract}


\date{\today}

\maketitle

{\small There is a popular summary in Appendix \ref{S-popular}.}

%=======================================================================
\section{Introduction}
\label{ch_INTRO}

Singularities play an interesting role in physics,
 and come in many different varieties,
 from the mathematically and philosophically challenging
 \cite{Curiel-1999ps,Curiel-2021sep,Earman-1996fp,Scheel-T-2014pu}
 to the more mundane \cite{Alawneh-K-1977siamr,Heckenberg-MSW-1992,Horsley-HMQ-2014sr}.
As the place where ``physics breaks down'' in a black-hole,
 we have the sense that anything might happen at a singularity.
This begs the question:
 are there things we might drop into a singularity
 that have fundamental properties that could be erased absolutely?
Crossing the event horizon, 
 for example,
 can lead to 
 the baryon number of matter falling in to a  black-hole
 not being conserved, 
 even if its mass-energy still persists \cite{Coleman-H-1993pla}.
Alternatively,
 the reverse scenario is also intriguing:
 we may intuitively have a sense that
 since the laws of physics have broken down,
 {anything} might \emph{emerge} from a singularity
 (see e.g. \cite[Chapter 3]{EarmanBCWS}).
Although perhaps most useful %easily dismissed as a
 as a plot device for science fiction stories,
 should we as concerned physicists nevertheless
 check what conservation laws
 might no longer hold?
But,
 if so,
 and given that immediately on leaving the singularity
 the laws of physics must then be followed,
 how could such artefacts
 manifest themselves?

In this article we consider the conservation of electric charge. 
In standard approaches to electromagnetism this is sacrosanct, 
 whether for
 local charge conservation (i.e. the differential version),
 or for 
 global charge conservation (the integral version).
Although local charge conservation is experimentally observed, 
 and is assumed here, 
 this 
 does not of itself guarantee
 %may not lead to 
 global charge conservation, 
 which
 instead arises via either of two mechanisms.
These mechanisms are:
 from local charge conservation
  and the assumption that the region of spacetime is topologically trivial, 
 or from the assumption that the excitation fields $\VD$ and $\VH$
  are real physical fields. 
In order to break global charge conservation
 it is necessary that both of these mechanisms
 no longer apply\footnote{If
   spacetime $\Mman$ has a non-zero third de Rham
  cohomology $H^3_{\textup{dR}(\Mman)}\ne0$,
  then there exists a closed 3--form current $\Je$ which is not exact, 
  i.e. $d \Je = 0$ but $\Je\ne d {\mathcal{H}}$
 for any excitation 2--form field ${\mathcal{H}}$. 
 Hence for 3--surfaces $\manU$ enclosing the temporary
  singularity $\int_\manU \Je\ne0$.}, 
 i.e.  we need
\begin{quote}
an extension to Maxwell's equations where $\VD$ and $\VH$ 
 are not fundamental physical fields, 
 but are merely gauge fields for the charge and current,
%% (we call this the \defn{electromagnetic condition})
\end{quote}
~~and 
\begin{quote}
a topologically non-trivial spacetime $\Mman$, 
 such as in this article,
 where we consider spacetimes with a temporary singularity. 
%% (we call the \defn{topological condition}).
\end{quote}

In \cite{Gratus-KM-2019foop-nocharge} we showed that
 a non trivial spacetime 
 can in principle break
 global charge conservation,
 but without detailing a specific set of electromagnetic constitutive relations
 that might support this. 
In  \cite{Gratus-MK-2020pra-area51} a  \emph{minimal} extension
 to Maxwell's equations was considered where $\VD$ and $\VH$ did not appear, 
 and which provided an additional axionic term 
 \cRed{to the vacuum} Maxwell's equations. 
Using this axionic term, 
 we now give in this article an 
 \emph{explicit construction} of an electromagnetic field configuration
 which can lead to the breaking of global charge conservation.

\cRed{In particular, 
 the classical Maxwell vaccum is augmented
 in quantum field theory to account for vacuum polarization
 for intense fields.
For example,
 the strong magnetic fields associated with magnetars
 induce non-trivial dielectric properties on vacuum \cite{Lai-H-2003aj},
 but longer established 
 alternatives
 for polarization of the vacuum 
 are (e.g.) the Euler-Heisenberg \cite{Heisenberg-E-1936zfp}
 or
 the Bopp-Podolski model
 \cite{Bopp-1940ap,Podolsky-1942pr,Gratus-PT-2015jpa}.
%There is some evidence, for example, that 
% \emph{...}. \\
However,
 in these,
 the model 
 corresponds to well defined $\VD$ and $\VH$, 
 in-effect also demanding they are measureable,
 when in fact this is not necessary
 \cite{Jackson-ClassicalED,Gratus-KM-2019ejp-dhfield,Heras-2011ajp,Roche-2000ajp,Roche-1998ejp,Landini-2014piers,Bork-1963ajp}.
To avoid this unnecessary assumption, 
 an alternative approach based \emph{purely} on the physically
 measurable fields $\VE$ and $\VB$
 was developed \cite{Gratus-MK-2020pra-area51}, 
 and this is what we use here.
}

The extension to Maxwell's equations \cite{Gratus-MK-2020pra-area51}
 considered the constitutive properties of a background medium
 on the basis that there may be an axion {\axiName} $\axi$
 that does not derive from an axion scalar field. 
\cRed{In contrast, 
 in}
 this article we posit that such an axion {\axiName}
 can exist independently in the vacuum,
 instead of just being a property of some background medium. 
We call this axion {\axiName} a topological axion
 because it is distinct from a standard axion
 %\cite{Tobar-MG-2019pdu,Visinelli-2013mpla},
 and because it has more interesting topological properties. 
In the context of electromagnetism, 
 axions are already
 an established area of research
 \cite{Carroll-FJ-1990prd,Colladay-K-1998prd,Itin-2004prd,Tobar-MG-2019pdu,Visinelli-2013mpla},
 usually arising in the context of an added coupling between
 the Maxwell fields and the field of an axion particle.
Such coupled Maxwell-axion dynamics
 allow situations where
 a background axion field can influence
 the behaviour of electromagnetic fields.
In the standard Maxwell theory,
 a \emph{piecewise} constant axion field for a medium
 \emph{is} detectable at boundaries \cite{Obukhov-H-2005pla}
 where the response can be equivalently cast as either
 a perfect electrical,
 or perfect magnetic conductor \cite{Lindell-S-2013ieeetap}.
Axionic responses were apparently observed experimentally
 \cite{Hehl-ORS-2008pla-relativisitic}
 via the magneto-electric effect
 in Cr$_2$O$_3$.
More recently axionic responses
 have also been proposed \cite{Li-WQZ-2010np-dynamical}
 and observed \cite{Wu-SKMO-2016s-quantised} in topological insulators.
Observations are,
 however,
 still controversial,
 with claims that evidence of violation
 of the so-called Post constraint \cite{Post-FSEM}
 can be explained by an admittance that describes surface states
 \cite{Lakhtakia-M-2015spie}.

In the domain of particle physics,
 axions have been proposed as candidates for dark matter
 \cite{Feng-2010araa-dark},
 but as yet no particle axions have been observed.
Given this context, 
 any self-consistent axionic model, 
 such as the one we propose here, 
 remains a valid candidate for exploration.

\begin{figure}
\centering
%\resizebox{0.900\columnwidth}{!}{\input{fig01-tsingularity.tex}}
\resizebox{0.900\columnwidth}{!}{\includegraphics{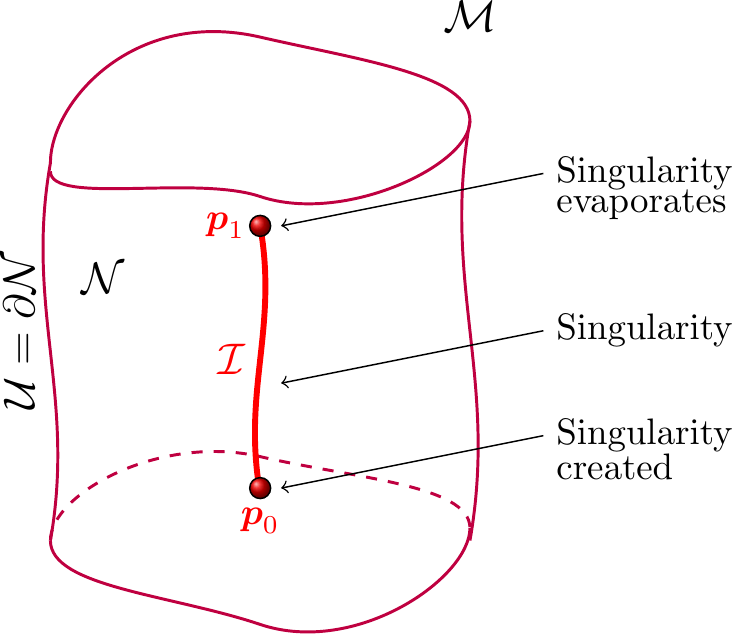}}
\caption{Here we show a temporary singularity. 
The region of spacetime $\manN$
 has a boundary $\manU=\partial \manN$ that encloses a singularity
 with a finite duration.
This might occur,
 for example,
 due to the formation at $\pini$ and subsequent evaporation at $\pfin$
 of a black-hole,
 which would first create and then remove
 a metric singularity in spacetime.
The temporary singularity $\interval$ is not part of $\Mman$.
}
\label{fig_singular}
\end{figure}

In this article we give the explicit construction
 for the simplest non trivial spacetime,
 namely Minkowski spacetime with a point removed,
  which provides us with a both a singularity and a non-trivial topology. 
However,
 since the machinery of transformation optics
 \cite{Dolin-1961ivuzr,Pendry-SS-2006sci,McCall-etal-2018jo-roadmapto}
 can be applied in a spacetime sense 
 \cite{McCall-FKB-2011jo,Kinsler-M-2015pnfa-tofu,Gratus-KMT-2016njp-stdisp},
 we also show that this apparently heavily restricted solution
 is also valid in the more general case of a temporary
 singularity,
 as depicted in figure \ref{fig_singular}. 
This may include cases
 where a black hole forms
 and then subsequently evaporates
 \cite{Coleman-H-1993pla,Abdolrahimi-PT-2019prd}.

% -  -  -  -  -  -

% -  -  -  -  -  -

In section \ref{ch_BH}
 we summarise the key points of a
 Maxwellian electrodynamics based on first-order operators
 \cite{Gratus-MK-2020pra-area51}
 that dispenses with $\VD$ and $\VH$
 and admits the topological 
 axion field. % $\axiTop$. 
In section \ref{ch_JTBH} we define a temporary
 singularity and its future. 
Next,
 in Section \ref{ch_SOLVE}
 we describe the key parts of the design of
 both the electromagnetic
 and the axionic fields,
 and and then in Section \ref{ch_EG} construct the solution
 in a way that conforms to the theoretical constraints.
Following this,
 in Section \ref{ch_TBH}
 we show how a transformation optics style approach
 allows us to show that our solution is not
 specific to the simple spacetime manifold
 in which we construct it,
 but is also valid for more realistic spacetime metrics,
 potentially
 even for the case of a forming then evaporating black-hole.
Finally,
 in Section \ref{ch_CON}
 we conclude.

%=======================================================================
\section{Topological Axions}\label{ch_BH}

{The topological axions we consider here 
 are allowed by a \emph{minimal} relaxation of Maxwell's equations
 where the role of the unmeasureable 
 electromagnetic excitation fields $\VD$ and $\VH$
 is explicitly demoted to that of mere
 gauge fields for the current.
As discussed in \cite{Gratus-MK-2020pra-area51}, 
 this extension permits a new axionic type of interaction
 with the electromagnetic field.
In addition to dispensing with $\VD$ and $\VH$,
 and in the presence of a spacetime singularity, 
 it has been shown that 
Maxwell's equations
 no longer need enforce global charge conservation
 \cite{Gratus-KM-2019foop-nocharge}; 
 but although
 posing some intriguing possibilities,
 this proposal did not include any 
 explicit examples of 
 constitutive relations,
 material configurations, 
 or field distributions.

Here, 
 by attributing this axonic interaction
 to the presence of a particle-like axion {\axiName}, 
 we find more general Maxwell's equations for vacuum.
Having freed the model from an explicit reference 
 to a material response as implied in \cite{Gratus-MK-2020pra-area51},
 we find that there is now}
 scope for an explicit solution that breaks
 global charge-conservation.
The starting point \cite{Gratus-MK-2020pra-area51}
 can be briefly summarized as follows.
Re-write Maxwell's equations as
%[
\begin{align}
  d\Fem
= 
  0 
\qquadand
  \Psi\VAct{\Fem}
=
  \Je,
\label{BH_CMCR}
\end{align}
%]
where\footnote{Here $\Gamma\Lambda^p\Mman$
   is the set of sections of the bundle $\Lambda^p\Mman$ of $p$--forms. 
  I.e. the statement $F\in\Gamma\Lambda^2\Mman$ means
   $F$ is a 2--form field on $\Mman$.
  However we usually just say $F$ is a 2--form,
   with the fact it is a field being implicit.}
 $\Fem \in \Gamma \Lambda^2 \Mman$ is the electromagnetic 2--form,
 $\Je \in \Gamma \Lambda^3 \Mman$ is the current 3--form,
 and $\Psi : \Gamma \Lambda^2 \Mman \to \Gamma \Lambda^3 \Mman$
 is a non-tensorial ``first order operator'' that satisfies
%[
\begin{align}
\begin{gathered}
  \Psi\VAct{\alpha_1+\alpha_2}
=
  \Psi\VAct{\alpha_1}+\Psi\VAct{\alpha_2},
\quad
  \Psi\VAct{\lambda\,\alpha}
=
 \lambda\Psi\VAct{\alpha}
,
\\
\text{and}\quad
  \Psi\VAct{f^2\,\alpha}
 -
  2f\,\Psi\VAct{f\,\alpha}
 +
  f^2\,\Psi\VAct{\alpha}
 =
  0
,
\label{BH_Psi_2nd_order}
\end{gathered}
\end{align}
%]
 for all $ \alpha, \alpha_1, \alpha_2 \in \Gamma \Lambda^2 \Mman$,
 $f \in \Gamma \Lambda^0 \Mman$
 and constants $\lambda\in\Real$.
The angle brackets $\VAct{...}$
  enclosing the arguments to $\Psi$
  are used
  to emphasise its non tensorial nature:
  notably we have that $\Psi \VAct{f \alpha} \neq f \Psi \VAct{\alpha}$.
This formulation
 permits a range of new possibilities\footnote{In terms of coordinates, 
  the general ${\Psi} \VAct{\Fem}$
 can be written using $i^a=g^{ab} i_{\partial_b}$,
 as
%[
\begin{align}
{\Psi} \VAct{\Fem} 
=
\left(\tfrac{1}{2} {\Psi}^{abc} \, \Fem_{bc}
 +
  \tfrac{1}{2} {\Psi}^{abcd} \left( \partial_b \Fem_{cd} \right)
\right)\star dx_a
,
\nonumber
%\hfill\textrm{F6a} 
%\label{FOO_CR_coord_alt}
\end{align}
%]
where
%[
\begin{align}
%~~ \textrm{where} ~
%\begin{aligned}
  {\Psi}^{abc}
=
  i^a \star \left( {\Psi} \VAct{dx^{b}\wedge dx^c} \right)
,
\nonumber
\\
~~ \textrm{and} ~
  {\Psi}^{abcd}
=
i^a\star  \left(
    {\Psi} \VAct{x^b \, dx^{c}\wedge dx^{d}}
   -
    x^b \, {\Psi} \VAct{dx^{c}\wedge dx^{d}}
  \right)
.
%\hfill\textrm{F6b}
\nonumber
%\end{aligned}
\label{GR_Coordfree_Comp}
\end{align}}, 
 but here we focus
 on the simplest ``axionic'' scheme.
This means we 
 replace the rather general \eqref{BH_CMCR} with
 an expression with a more familiar appearance, 
 namely %(e.g. compare with \cite{Itin-2004prd})}
%[
\begin{align}
  d\Fem = 0 
\qquadand
  d\star(\kappa(\Fem))
 +
  \axiTop\wedge \Fem 
=
  \Je
.
%\label{BH_gen_Max}
\end{align}
%]
Here $\kappa$ is the usual constitutive tensor
 for a medium,
 and $\axiTop$ is a 1--form field
 generating an additional axion-like interaction.
By setting
 $\Psi\VAct{\Fem}= d\star(\kappa(\Fem)) + \axiTop\wedge \Fem$
 we see that \eqref{GR_Coordfree_Comp}
 is an example of \eqref{BH_Psi_2nd_order}
 since
   %[
\begin{align*}
  \Psi&\VAct{f^2\Fem} - 2f\,\Psi\VAct{f\,\Fem} + f^2\,\Psi\VAct{\Fem}
  \qquad
\\
&=
  d\star(\kappa(f^2\,\Fem))
 +
  \axiTop\wedge (f^2\,\Fem)
 -
  2 f\, d\star(\kappa(f\,\Fem)) 
\\
&\quad
 -
  2 f\,\axiTop\wedge (f\,\Fem)
 +
  f^2\,d\star(\kappa(\Fem))
 +
  f^2\,\axiTop\wedge \Fem
\\
&=
  d\left(f^2\star(\kappa(\Fem))\right)
 -
  2 f\, d\left(f\star(\kappa(\Fem))\right)
 +
  f^2\,d\star(\kappa(\Fem))
\\
&=
  2 f \, d f\wedge \star(\kappa(\Fem))
 +
  f^2 \, d\star(\kappa(\Fem))
 -
  2 f \, df\wedge \star(\kappa(\Fem))
\\
&\quad
 - 
  2 f^2 \, d\star(\kappa(\Fem))
%   \\&\quad
   + f^2 \, d\star(\kappa(\Fem))
=
  0
.
\end{align*}
   %]
In a vacuum
 (i.e. with a trivial $\kappa$),
 \cRed{we find that adding this axionic interaction 
 naturally adapts}
 the combined Maxwell-Amp{\`e}re-Gauss equation ($\Psi\VAct{\Fem}=J$)
 into
%[
\begin{align}
  {\Psi} \VAct{\Fem}
= 
  d \star \Fem + \axiTop \wedge \Fem
=
  \Je
.
\label{BH_dstarF}
\end{align}
%]
In this formalism,
 the 2--form $\Fem$ is untwisted,
 while the $\axiTop$ and $\Je$ are twisted.
We emphasise that \eqref{BH_dstarF}
 proposes an extension to the usual vacuum Maxwell equation 
 ($d \star \Fem = \Je$) 
 in which there is no role 
 (or even sensible definition) for
 the unmeasurable fields $\VD$ and $\VH$.

We denote the axion {\axiName} $\axiTop$ 
 a ``topological axion''
 \cRed{for two reasons:
 because it has topological consequences
 distinct from 
 those axions typically used
 in models of axion--electromagnetism interaction, 
 and because of the similarities with the standard proposal for 
 the still hypothetical axion particle.}
%We call this {\axiName} We can
% compare and contrast the $\axiTop$ model
% with the standard proposal for the nature of
% the (still hypothetical) axion particle.

The standard axions are given by
 a twisted scalar field $\StanAxi\in\Gamma\Lambda^0\Mman$,
 and
 their coupling with the electromagnetic field in vacuum
 \cRed{defined in the Lagragian by
  $\tfrac12 \StanAxi dA \wedge dA$ \cite[eqn.(2)]{Tobar-MG-2019pdu},}
 is given by \cite[eqns.(3,4)]{Tobar-MG-2019pdu}
%[
\begin{align}
  d\star \Fem
 +
  d\StanAxi\wedge \Fem
=
  \Je
.
\label{BH_dstarF_stdAx}
\end{align}
%]
\cRed{The similarity between \eqref{BH_dstarF}
 and \eqref{BH_dstarF_stdAx}
 demonstrates that both interactions are indeed of the same 
 ``axionic'' type.}

We {could} therefore obtain \eqref{BH_dstarF}
 by insisting that $\StanAxizeta$ is exact,
 i.e. $d\StanAxi=\StanAxizeta$.
However,
 it is then {no longer} possible
 for a field $\StanAxiSource$
 to act as a \emph{source}
 for topological axions,
 since this would imply $\StanAxiSource=d\StanAxizeta=d^2\StanAxi=0$.
We show below that
 %(cf. Eqs. \eqref{BH_axi_Source} and \eqref{BH_zetaS_F_abhor})
 to break global charge conservation requires topological axions,
 for which \cRed{there is an axion {\axiNameSource} $\axiSource$, 
 where}
 %$\axiSource = d \axiTop \ne 0$. 
\begin{align}
  \axiSource = d \axiTop \ne 0
\label{eqn-Zdaxion}
\end{align}
 This is consistant with local charge conservation ($d\Je=0$)
 provided $\Fem$ and $\axiSource$ satisfy
 the constraint $\axiSource\wedge\Fem=0$, 
 as is evident by taking the exterior derivative of \eqref{BH_dstarF}.
The topological axions utilised here
 are therefore distinct from the standard axion hypothesis,
 wherein $d\StanAxi=\StanAxizeta$.
Nevertheless,
 since even standard axions have not been detected yet,
 a widening of axionic theory to include such topological axions
 is not ruled out by any experimental results to date.

Here 
   we are not directly concerned
   with the dynamics of $\axiTop$ or $\axiSource$, 
 but it is nevertheless an interesting consideration.
For a massless topological axion, 
 we can
 define an axion source term $\axiTrueSource  \in\Gamma\Lambda^3\Mman$, 
 and set
\begin{align}
   \axiTrueSource = d \star \axiSource
 \label{eqn-xiequalsdZ}
\end{align}
 so that
  we could if desired replace \eqref{eqn-Zdaxion}
   with the dynamical equation
\begin{align}
  d \star d \axiTop = \axiTrueSource
.
\end{align}
In what follows we specify the axion {\axiName}
 in terms of the axion {\axiNameSource} $\axiSource$, 
 but this could be converted into a specification
 of $\axiTrueSource$ using \eqref{eqn-xiequalsdZ} if desired.

Notwithstanding any wider aspects of the field dynamics
 and the possibility of a Lagrangian (see Appendix \ref{ch_Lagrangian}), 
 neither of which are needed for the work herein,
 we now proceed to consider the primary requirements for our results.

%=======================================================================
\section{{The future region of a temporary singularity}}\label{ch_JTBH}

Before discussing how global charge conservation
 can be violated in a spacetime supporting topological axions, 
 we would like to clarify some aspects of the singularity
 and its surrounding manifold that together
 provide the backcloth of our construction.

% -  -  -  -  -  -

\begin{figure}
%\resizebox{0.500\columnwidth}{!}{\input{fig02-tpenrose.tex}}
\resizebox{0.500\columnwidth}{!}{\includegraphics[angle=90]{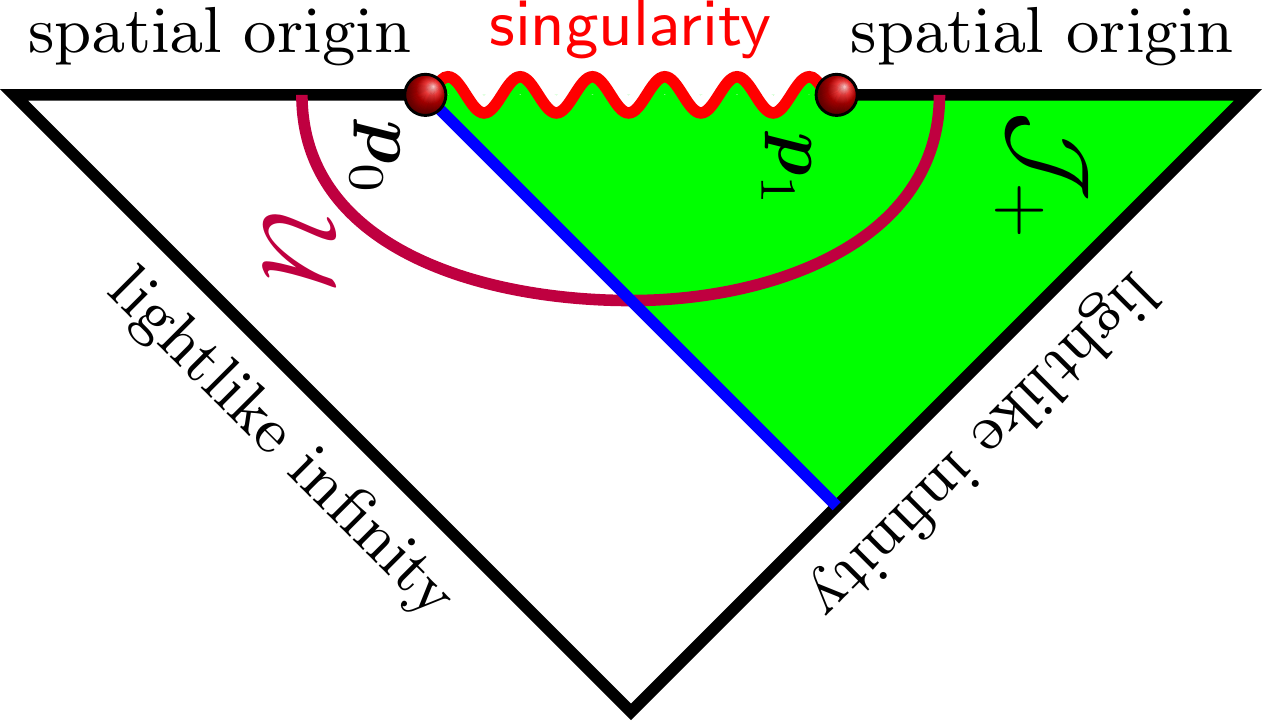}}
\caption{Penrose diagram
 for a temporary timelike singularity,
 at the origin,
 between the points $\pini$ and $\pfin$. 
Since light rays travel at 45$^\circ$ on these diagrams, 
 the future of the singularity 
 (its ``causal cone'')
 consists of both the light cone or null cone (blue line), 
 and the {\colInCone} region within it.
This Penrose diagram contains essentially the same information 
 as the more general view given in figure \ref{fig_singular}, 
 but with its lightlike infinity boundaries
 mapped nonlinearly down to a finite extent.
With time passing vertically upwards, 
 the radial distance extends horizontally rightwards,
 away from the time axis.
The {\colUsurf} line denoted $\manU$
 is a 3--sphere, 
 and the region inside $\manU$
 includes the singularity.
}
\label{fig_Penrose_time-sing}
\end{figure}

% -  -  -  -  -  -

\begin{figure}
%\resizebox{0.700\columnwidth}{!}{\input{fig03-spenrose.tex}}
\resizebox{0.700\columnwidth}{!}{\includegraphics{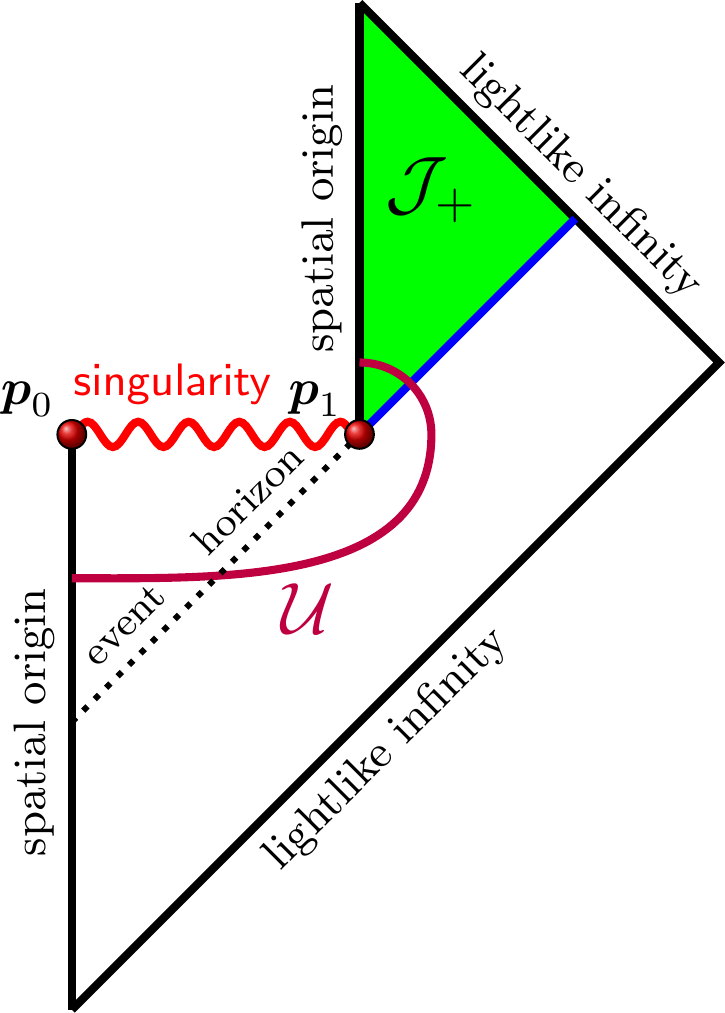}}
\caption{Penrose diagram for a temporary spatial singularity, 
 perhaps due to a forming then evaporating black-hole.
Using the same conventions as figure \ref{fig_Penrose_time-sing}, 
 here the spacelike singularity is instead extended horizontally, 
 thus creating the event horizon,
 and giving the correct causal structure.
The {\colUsurf} line is a
  3--sphere $\manU$, 
 and
 the region inside $\manU$ includes the singularity.
}
\label{fig_Penrose}
\end{figure}

% -  -  -  -  -  -

\begin{figure}
\centering
%\resizebox{0.950\columnwidth}{!}{\input{fig04-singularityevent.tex}}
\resizebox{0.950\columnwidth}{!}{\includegraphics{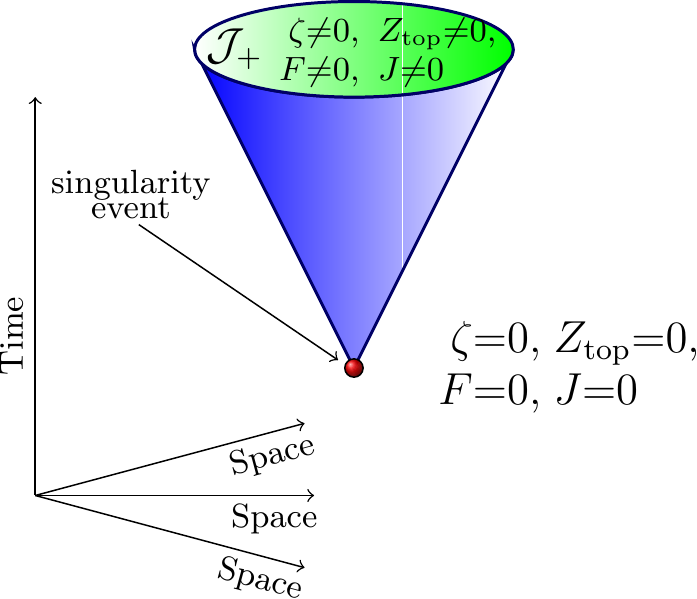}}
\caption{A ``singularity event'' is excised from spacetime.
The causal cone of this singularity
 is outlined by the light cone ({\colLiCone})
 and its interior ({\colInCone}).
The resulting non-trivial topology,
 in concert with the
 demotion of $\VD$ and $\VH$
  to mere gauge fields,
 allows global charge conservation to be broken.
Our analytic solution achieves this.
In the region 
 inside the causal cone the fields vary in time and space.
Outside the causal cone they are everywhere zero.
}
\label{fig_CQ}

\end{figure}

% -  -  -  -  -  -

Specifying whether a spacetime has a temporary singularity
 is subtle since the singularity itself is not part of the spacetime. 
There are a plethora of definitions
 regarding singularities and completeness \cite{Curiel-1999ps,Earman-1996fp}.
The approach taken here is to state that
 a spacetime $\Mman$ has a \defn{temporary singularity}
 if there exists a topological 3--sphere $\manU\subset\Mman$
 which is \emph{not} the boundary of a (compact) 4--dimensional ball. 
We see for example on figure \ref{fig_singular}
 that $\manU=\partial\manN$, 
 but $\manN$ is not a 4--dimensional ball as it has a hole in it. 
The 3--sphere $\manU$ separates $\Mman$ into two regions
 which can be called the ``inside'' and ``outside''.
The inside region is the one that 
 has the temporary singularity, 
 and the outside region goes off to infinity, 
 as depicted on the Penrose diagram \cite{MTW} 
 on figure \ref{fig_Penrose_time-sing}.
In what follows we assume there 
 is precisely one temporary singularity,
 i.e. inside the 3--sphere $\manU$
 there do not exist two or more 3--spheres inside each of which
 is a temporary singularity, 
 although
 it is easy to extend the analysis. 
Also, 
 if we were to consider 
 the case of a temporary black hole, 
 then in addition to containing a temporary singularity, 
 the spacetime would also require the 
 properties of the metric to be specified, 
 i.e. the existence of an event horizon 
 (see figure \ref{fig_Penrose}),
 and the metric becoming singular
 as the singularity is approached.

Of course there are many such 3--spheres
 which surround the temporary singularity
 and we exploit this to define the future $\FutLC$ of the singularity. 
We say that a point $p\in\FutLC$ if for all 3--spheres $\manU$ 
 surrounding the singularity and with $p$ outside $\manU$
 then there exists a causal curve (timelike or lightlike) which passes from
 $\manU$ to $p$.  
Conversely,
 we say that $p\not\in\FutLC$ if there exists a 3--sphere $\manU$
 surrounding the singularity,
 with $p$ on the outside of $\manU$, 
 which does not intersect the backward causal cone of $p$.

As an example,
 let $\Mman=\Real^4\backslash\Set{0}$ be Minkowski space excluding the origin,
 then
%[
\begin{align}
  \FutLC
=
  \Set{(t,x,y,z) \, \Big| \, t \ge \sqrt{x^2+y^2+z^2}}
.
\label{BH_def_FutLC}
\end{align}
%]
 represents the future of the excised origin as shown in figure \ref{fig_CQ}. 
This example will be used in section \ref{ch_EG} below
 to construct a combined electromagnetic and axion field configuration
 which breaks global charge conservation.

A more general example of a temporary singularity
 can be constructed by considering a spacetime $\MmanSup$
 and then excluding a compact set $\interval\subset\MmanSup$,
 so that $\Mman = \MmanSup \backslash \interval$. 
Any 3--sphere surrounding $\interval$ 
 is not the boundary of a compact 4--dimensional ball. 
An example of $\interval$ is
 a finite interval with end points $\pini,\pfin\in\interval$
 as shown in  Fig. \ref{fig_singular}.
Let $\FutLC^{\text{sup}}(\interval) \subset \MmanSup$
 be the future causal cone of $\interval$. 
Then $\FutLC^{\text{sup}}(\interval) \backslash \interval \subset \Mman$
 and
%[
\begin{align}
  \FutLC
=
  \FutLC^{\text{sup}}(\interval)\backslash\interval
\label{BH_Fut_MmanSup}
\end{align}
%]
This is shown in appendix \ref{ch_Proofs}. 
Thus the definition of $\FutLC$
 coincides with the usual definition
 of the future causal cone for $\interval$.

However, 
 even this more general construction is still
 not the same as that for a temporary black hole.
In section \ref{ch_TBH} we will link the above examples
 using the techniques of transformation optics,
 in order to show that our conclusions based on a point singularity
 are in fact significantly more general.

%=======================================================================
\section{Technical ingredients for global charge conservation violation}\label{ch_SOLVE}

We can now state precisely the technical ingredients
 leading to our solution that manifests 
 non-conservation of global charge.
Since we only consider topological axions, 
 from now on we
 write the axion {\axiName} $\axiTop$ as just $\axi$.
In order to break global charge conservation,
 we require a spacetime $\Mman$ with a temporary singularity,
 with the future causal cone $\FutLC$ as described in section \ref{ch_JTBH},
 and the model of electromagnetism described in section \ref{ch_BH},
 with its topological axions, i.e.
%   STYLE: we are unfortunately stuck with this reversed section ordering 
%      unless we split this sentence over the following enumerated list
~
\begin{enumerate}[label=(\roman*)]

\item \label{enum-solution-F}
 An electromagnetic field $\Fem\in\Gamma\Lambda^2\Mman$ with
  support only in $\FutLC$.

\item \label{enum-solution-J}
 An electric current
 $\Je\in\Gamma\Lambda^3\Mman$ with support
  only in $\FutLC$.

\item \label{enum-solution-Z}
 An \emph{axion} current $\axi \in \Gamma \Lambda^1 \Mman$
  with support
  only in $\FutLC$; and
 an \emph{axion} source $\axiSource\in\Gamma\Lambda^2\Mman$,
 which also has support only in $\FutLC$.

\end{enumerate}

The fields $\Fem$,
 the electric current $\Je$,
 the axion {\axiName} $\axi$,
 and the axion {\axiNameSource} $\axiSource$
 all satisfy the following criteria:
 %\ref{enum-criteria-Maxion} to \ref{enum-criteria-Z}:

\begin{enumerate}[label=(\alph*)]

\item
The vacuum Maxwell axion relations \eqref{BH_dstarF}.
  \label{enum-criteria-Maxion}

\item
The monopole free condition
%[
\begin{align}
d\Fem = 0\,.
\label{BH_dF}
\end{align}
%]
  \label{enum-criteria-dF}

\item
The {local} conservation of charge
%[
\begin{align}
d\Je=0\,.
\label{BH_consv_Charge}
\end{align}
%]
  \label{enum-criteria-dJ}

\item
The axion {\axiName} has
 a a {source}{-like} {\axiNameSource} {$\axiSource$}
 where 
%[
\begin{align}
d\axi = \axiSource\,.
\label{BH_axi_Source}
\end{align}
%]
  \label{enum-criteria-Z}

%\end{enumerate}

\noindent\hspace{-29pt} % hack to fake a no-indent whilst remaining in the \enumerate.
{The final and most important criterion that
 we require
 \ref{enum-solution-F} to \ref{enum-solution-Z}
 satisfy is}\footnote{\cRed{Note 
  that there is no axion property
  (i.e. neither $\axi$ nor $\axiSource$)
  that contributes directly to the current in Maxwell equations;
  these axions carry no electric charge.
 There is therefore a clear distinction
  between the nature of the conventional current $\Je$
  and the axionic interaction term $\axi \wedge \Fem$.
 Nevertheless, 
  it is interesting to note that \emph{if} one
  were to claim $\axi \wedge \Fem$ as part
  of some new augmented current
  $\Je_\textup{new} = \Je - \axi \wedge \Fem$, 
  this $\Je_\textup{new}$ \emph{would} be automatically conserved
  both globally and locally, 
  since $d \star \Fem = \Je_\textup{new}$.}}

%\begin{itemize}

\item \label{enum-criteria-Qe} % label ref does not work 
 The total charge is \emph{not} {globally} conserved.

\end{enumerate}%\end{itemize}

\noindent
Obviously,
 \emph{outside} the future causal cone $\FutLC$,
 the current $\Je=0$;
 and for any spatial hypersurface $\SigmaPast$
 which does not intersect the $\FutLC$ we have
%[
\begin{align}
\int_{\SigmaPast} \Je = 0.
\label{BH_past_current_int}
\end{align}
%]
Therefore,
 to show that the last criterion \ref{enum-criteria-Qe}
 is satisfied
 we need %the total charge after the origin to be non-zero,
 %so 
 that for a spatial hypersurface $\SigmaFut$
 which intersects $\FutLC$
%[
\begin{align}
\QFut=
\int_{\SigmaFut} \Je
\ne 0
.
\label{BH_future_current_int}
\end{align}
%]

\begin{figure}
\centering
%\resizebox{0.950\columnwidth}{!}{\input{fig05-chargeintegral.tex}}
\resizebox{0.950\columnwidth}{!}{\includegraphics{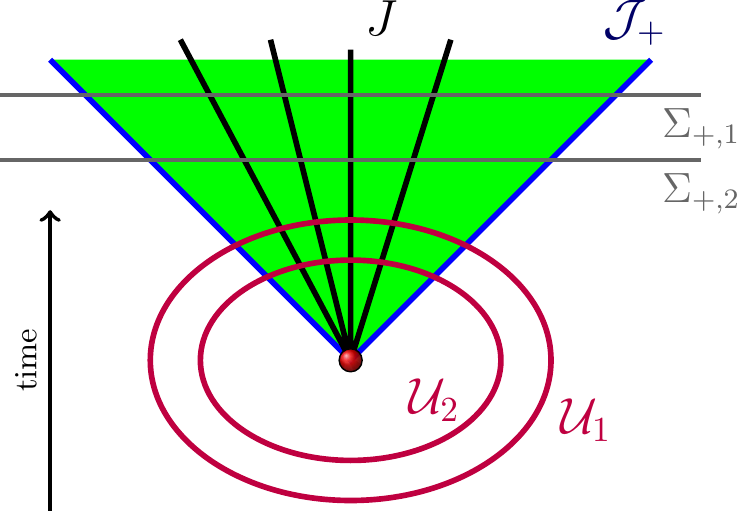}}
\caption{Charge integral, 
 as per \eqref{BH_future_current_int} and \eqref{BH_current_int_U}; 
 using the representation of forms as lines and surfaces
 given in appendix \ref{ch_Supp}.
The conserved current $J$
 can be represented as worldlines (black), 
 and
 the integral of the form over a hypersurface $\Sigma_{+,i}$
 is given by the number of lines that cross it --
 one  may think of it as counting the number of charge at that time. 
Since there are no charges outside $\FutLC$ (bounded by {\colLiCone} lines)
 then clearly
 $\QFut=\int_{\manU_1}J=\int_{\manU_2}J=\int_{\Sigma_{+,1}}J=\int_{\Sigma_{+,2}}J$.}
\label{fig_conserveJ}
\end{figure}

``Away from'' the temporary singularity, 
 i.e. if there is a
  topological 3--sphere which surrounds the temporary singularity,
  but which does not intersect 
 ${\SigmaFut}$ (the region under consideration), 
 local charge conservation \eqref{BH_consv_Charge}
 implies global charge conservation.
Hence $\QFut$ is independent of $\SigmaFut$ as long as
  $\SigmaFut$ is away from the temporary singularity.
Thus for
  any topological 3--sphere $\manU$ surrounding the temporary
  singularity we deduce (cf. figure \ref{fig_conserveJ})
%[
\begin{align}
  \QFut
=
  \int_\manU \Je
\label{BH_current_int_U}.
\end{align}
%]

% -  -  -  -  -  -

% -  -  -  -  -  -

The axion {\axiNameSource} $\axiSource$ is
 constrained by the electromagnetic field,
 and
 from \eqref{BH_dstarF},
 \eqref{BH_dF},
 \eqref{BH_consv_Charge} and \eqref{BH_axi_Source} we have
%[
\begin{align}
\axiSource\wedge\Fem=0.
\label{BH_zetaS_F_abhor}
\end{align}
%]
One way to
 satisfy this constraint
 is to have an axion {\axiNameSource}
 entirely separate from
 the electromagnetic field,
 i.e. for the supports of  $\axiSource$ and $\Fem$ to be disjoint.
We might therefore think of $\axiSource$
 as being somewhat analogous to a type-I superconductor
 that expels magnetic fields.
However,
 this {complete separation could be relaxed}
 if we wished to give $\axiSource$ more freedom.

\def\dual#1{{\widetilde{#1}}}

Notably, 
 if we 
 perform the 3+1 splits of $F$ and $\axiSource$ 
 with respect to a field of observers $V\in\Gamma T\Mman$, $g(V,V)=-1$,
 to give
%[
\begin{align}
\begin{gathered}
  \Fem
=
  E\wedge \dual{V}
 +
  \star \left( B \wedge\dual{V}\right)
\\
\textrm{and}\qquad
  \axiSource
=
  \axiSE \wedge \dual{V}
 +
  \star \left( \axiSB  \wedge \dual{V}\right)
\end{gathered}
\label{BH_EM_zeta_3+1split}
\end{align}
%]
where $\dual{V}=g(V,-)\in\Gamma\Lambda^1\Mman$ 
is the metric dual of $V$.
Then \eqref{BH_zetaS_F_abhor} becomes
%[
\begin{align}
  g(E,\axiSB) 
 +
  g(B,\axiSE)
=
  0
\label{BH_zetaS_F_abhor_31}
\end{align}
%]
where $g$ is the spacetime metric
 (see proof in Appendix \ref{ch_MetricProof}). 
With respect to the observer $V$ the spatial parts of $\axiSource$
 are observed to be $(\axiSE,\axiSB)$, 
 analogous to the way
 that the spatial parts of $F$ are $(E,B)$. 
Equation  \eqref{BH_zetaS_F_abhor_31} implies that
 to conserve charge locally we only
 need a single constraint on the polarisations of 
 $\VE, \VB, \axiSE, \axiSB$.
Note that below we will ensure this condition holds
 for our field solution
 by specifying the supports of $\Fem$ and $\axiSource$
 to be disjoint.
This specification is both convenient and simplifying, 
 although not necessary.

%=======================================================================
\section{Construction of $\axi$ and $\Fem$}\label{ch_EG}

We now proceed
 to define explicit forms for
 $\axi$ and $\Fem$
 that satisfy
 criteria \ref{enum-criteria-Maxion}--\ref{enum-criteria-Qe} above. %(a)-(e)
In what follows, 
 for convenience 
 we specify particular sizes --
 $\tfrac{3}{10}$, $\tfrac{5}{10}$, 
 and so on --
 but any other convenient sizing that matches the requirements is 
 equally useful.

\begin{figure}
%\resizebox{0.900\columnwidth}{!}{\input{fig06-bumpfnr-a.tex}}
%\resizebox{0.900\columnwidth}{!}{\input{fig06-bumpfnz-b.tex}}
%\resizebox{0.900\columnwidth}{!}{\input{fig06-bumpfnrho-c.tex}}
\resizebox{0.900\columnwidth}{!}{\includegraphics{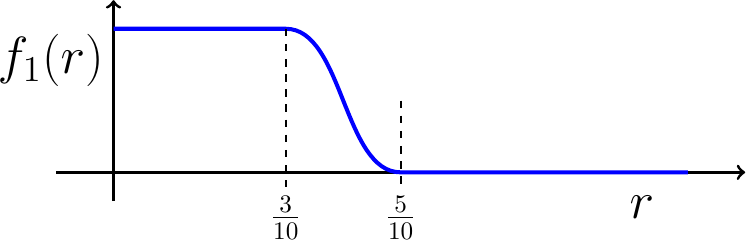}}
\resizebox{0.900\columnwidth}{!}{\includegraphics{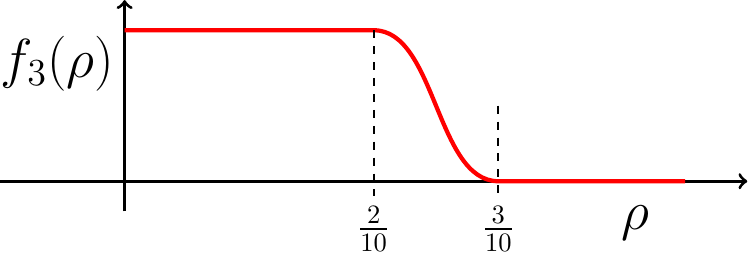}}
\resizebox{0.900\columnwidth}{!}{\includegraphics{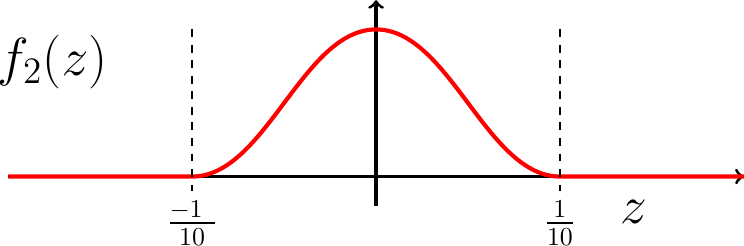}}
\caption{The three smooth ``bump functions''
 used to construct our axisymmetric field solutions.
There is one bump function defined for each of the radial $r$ coordinate, 
 the longitudinal $z$, 
 and the ``torus-radial'' $\rho$:
 i.e. 
 $\funr(r)$,
 $\funz(z)$,
 and
 $\funA(\rho)$.}
\label{fig_three_funs}
\end{figure}

Note that
 outside the causal cone $\FutLC$, 
 $\Fem$ and $\axi$ are both zero,
 and to ensure the solution remains well behaved
 we construct them
 using
 bump functions.
For this construction,                              % \cRed{
 $\Mman$ is Minkowski spacetime excluding the origin
 and $\FutLC$ is given by \eqref{BH_def_FutLC}.     % }
We will use three bump functions,
 denoted
 $\funr(r)$,
 $\funz(z)$, 
 and $\funA(\rho)$.
They
 are smooth,
 non-negative functions
 of the type shown in figure \ref{fig_three_funs}.
Although the functions must have properties that obey
 specific conditions, 
 they are otherwise arbitrary.
The conditions are
%[
\begin{equation}
\begin{aligned}
\funr(r) =
\begin{cases}
1 & r<\tfrac{3}{10},
\\
\text{strictly decreasing} & \tfrac{3}{10} < r < \tfrac{5}{10},
\\
0 & r>\tfrac{5}{10},
\end{cases}
\end{aligned}
\label{EG_def_f1}
\end{equation}
%]
and
%[
\begin{equation}
\begin{aligned}
\funz(z) =
\begin{cases}
\text{positive for}
&
\norm{z}<\tfrac{1}{10},
\\
0
\quadtext{for}
&
\norm{z}>\tfrac{1}{10},
\end{cases}
\end{aligned}
\label{EG_def_f2}
\end{equation}
%]
and
%[
\begin{equation}
\begin{aligned}
\funA(\rho)
=
\begin{cases}
\text{1} & \rho<\tfrac{2}{10},
\\
\text{strictly decreasing} & \tfrac{2}{10} < \rho < \tfrac{3}{10},
\\
0 & \rho>\tfrac{3}{10}.
\end{cases}
\end{aligned}
\label{EG_def_f3}
\end{equation}
%]
%See figure \ref{fig_three_funs}.

\def\tallthing{\vphantom{T}}

% -  -  -  -  -  -

% -  -  -  -  -  -

\emph{First,} 
 note that 
 the above functions
 can be combined to ensure that the the physical axion field
 $\axi$ 
 is non-zero only in the regions where required
 (in cylindrical polars $(t,r,\theta,z)$,
  and with $t>0$),
 i.e.
%[
\begin{align}
  \axi
=
  \funr\left(\frac{r}{t}\right)
  ~
  \funz\left(\frac{z}{t}\right)
  ~
  d\left(\frac{z}{t}\right)
,
\label{EG_def_axi}
\end{align}
%]
so that
%[
\begin{align}
  \axiSource
=
  d\axi
=
  \funr'\left(\frac{r}{t}\right)
  ~
  \funz\left(\frac{z}{t}\right)
  ~
  d\left(\frac{r}{t}\right)
  \wedge
  d\left(\frac{z}{t}\right)
,
\label{EG_axiSource}
\end{align}
%]
 where from \eqref{EG_def_f1}, \eqref{EG_def_f2}, and \eqref{EG_axiSource},
 the support of $\axiSource$ is in the region when
 $\norm{z/t}
 \le \tfrac{1}{10}$ and $\tfrac{3}{10} \le r/t \le \tfrac{5}{10}$.
\cRed{This arrangement of $\axiSource$
 could be derived from and attributed to its related source $\axiTrueSource$.
Note that 
 from \eqref{eqn-xiequalsdZ},
 the support of $\axiTrueSource$ is contained within
 that of $\axiSource$, 
 i.e. $\textrm{supp}(\axiTrueSource) \subseteq \textrm{supp}(\axiSource)$.}

\emph{Second,}
 we
 let the electromagnetic potential $\Aem$,
 which will define $\Fem$,
 be given by
%[
\begin{align}
  \Aem
=
  \funA\left(\frac{\rho}{t}\right) 
  ~
  d\theta
,
\label{EG_def_A}
\end{align}
%]
where
%[
\begin{align}
  \rho^2 
=
  z^2
 +
  \left( r - \tfrac{4}{10} \right)^2
.
\label{EG_def_rho}
\end{align}
%]
so that
%[
\begin{equation}
\begin{aligned}
  & \Fem
= dA
=
  \funA' \left(\frac{\rho}{t}\right)
  ~
  d\left(\frac{\rho}{t}\right)
  \wedge
  d\theta
\\
&=
  \funA'  \!
    \left(\frac{\rho}{t}\right)
  \frac{1}{t^2}
  \left(
    t \, d\rho\wedge d\theta 
   -
    \rho\,dt\wedge d\theta
  \right)
\\
&=
  \funA' \!
    \left( \frac{\rho}{t} \right)
   \!
  \frac{1}{t^2\rho}
  \left[
    t z \, dz \wedge d\theta
  + t\,(r-\tfrac{4}{10}) 
    dr \wedge d\theta
  - \rho^2 dt \wedge d\theta
  \right]
\end{aligned}
\label{EG_F}
\end{equation}
%]
From \eqref{EG_F} we see that
 the support of $\Fem$
 is when $\funA'(\rho/t)\ne 0$,
 i.e. when $\tfrac{2}{10}\le \rho/t \le \tfrac{3}{10}$.

% -  -  -  -  -  -

\begin{figure}
%\resizebox{0.900\columnwidth}{!}{\input{fig07-support2D.tex}}
\resizebox{0.900\columnwidth}{!}{\includegraphics{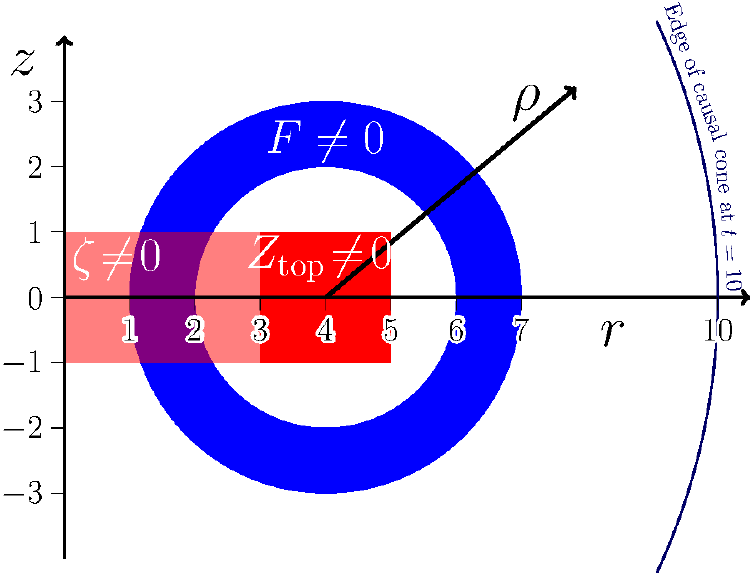}}
\caption{The support of the
 electromagnetic field  $\Fem$ (blue annulus),
 the axion {\axiName} $\axi$ (light red and dark red blocks),
 and
 the axion {\axiNameSource} $\axiSource$ (dark red block),
 at the moment when $t=10$. 
The region where $\axi\wedge F\ne 0$
 is given by the intersection of the blue and light red regions.  
The figure is rotated about the $z$--axis,
 as seen in figure \ref{fig_supp_F_axi_z0}.
This means that the support of $\Fem$ (blue)
 becomes a solid hollow torus 
  (as in figures \ref{fig_3d_BlueBit} and \ref{fig_3d_BothBits});
 the support of $\axiSource$ (dark red)
 becomes a solid torus with a square cross section
  (as in figures \ref{fig_3d_RedBit} and \ref{fig_3d_BothBits});
 and the support of $\axi$ (light and dark red)
  becomes a thick disc (as in figure \ref{fig_3d_RedPinkBit}).}
\label{fig_supp_lines_F_axi}
\end{figure}

% -  -  -  -  -  -

\begin{figure}
\centering
\resizebox{0.90\columnwidth}{!}{\includegraphics{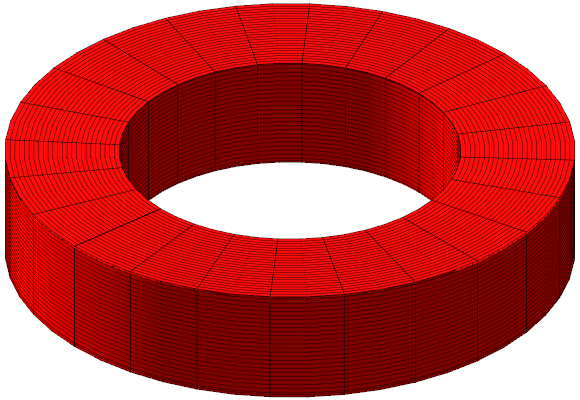}}
\caption{The support of the axion {\axiNameSource} $\axiSource$ (dark red), 
 at $t=10$, as a 3d plot.}
\label{fig_3d_RedBit}
\end{figure}

% -  -  -  -  -  -

\begin{figure}
\centering
\resizebox{0.90\columnwidth}{!}{\includegraphics{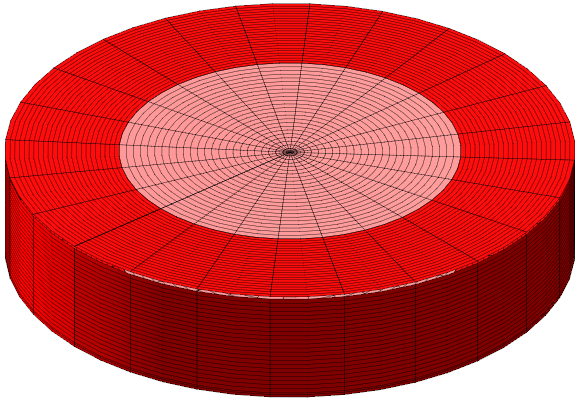}}
\caption{The support of the axion {\axiName} $\axi$
  (both light red and dark red parts)
 and the axion {\axiNameSource} $\axiSource$ (dark red only), at $t=10$, as a 3d plot.}
\label{fig_3d_RedPinkBit}
\end{figure}

% -  -  -  -  -  -

\begin{figure}
\centering
\resizebox{0.90\columnwidth}{!}{\includegraphics{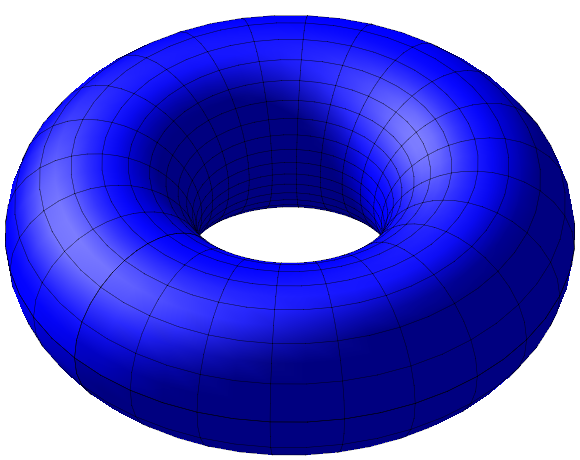}}
\caption{The support of the electromagnetic field $\Fem$ (blue), 
at $t=10$, as a 3d plot.
According to its definition in \eqref{EG_F},
 and as can be seen from figure \ref{fig_3d_BothBits}, 
 this is not solid, 
 but is a torus-shaped thick but hollow shell.}
\label{fig_3d_BlueBit}
\end{figure}

% -  -  -  -  -  -

\begin{figure}[t]
\resizebox{0.90\columnwidth}{!}{\includegraphics{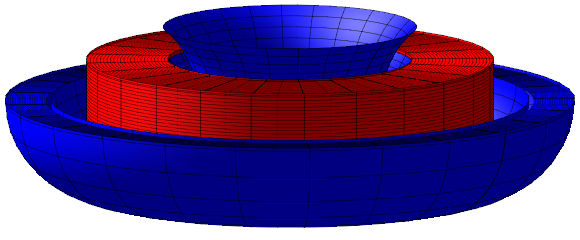}}
\resizebox{0.90\columnwidth}{!}{\includegraphics{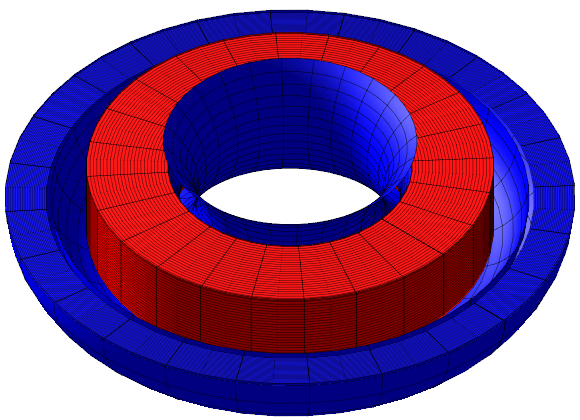}}
\caption{The combined 
 support of the axion {\axiNameSource} $\axiSource$ (red)
  and a cut of the support of the electromagnetic field $\Fem$ (blue), 
 shown in two views.}
\label{fig_3d_BothBits}
\end{figure}

% -  -  -  -  -  -

\begin{figure}
%\resizebox{0.900\columnwidth}{!}{\input{fig12-altsupport2D.tex}}
\resizebox{0.990\columnwidth}{!}{\includegraphics{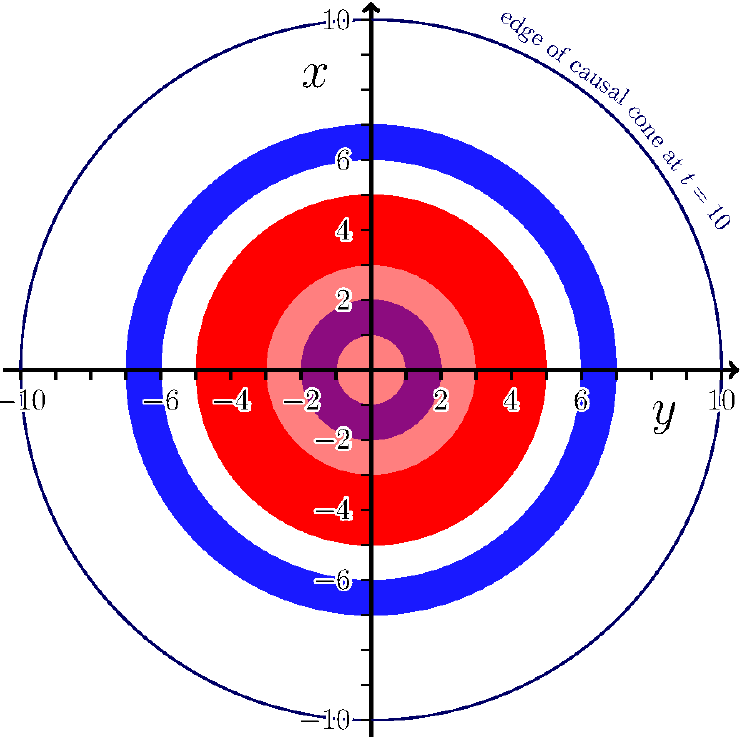}}
\caption{An alternative view of the supports of the electromagnetic field $\Fem$ (blue),
 the axion {\axiName} $\axi$ (light and dark red),
 and the axion {\axiNameSource} $\axiSource$ (dark red),
 at the moment when $t=10$ on the slice $z=0$.}
\label{fig_supp_F_axi_z0}
\end{figure}

% -  -  -  -  -  -

The specifications \eqref{EG_def_axi} and \eqref{EG_def_A}
 mean that
 for $\left. r^2 + z^2 \right. > t^2$,
 both $\Fem=0$ and $\axi=0$;
 thus the support of
 these lie inside the causal cone $\FutLC$.
Having now defined $\axi$ and $\Fem$,
 we can also define $\Je$ using \eqref{BH_dstarF}.
Further,
 since the supports of $\axiSource$ and $\Fem$ are disjoint, 
 \eqref{BH_zetaS_F_abhor} is satisfied, 
 which implies, 
 by taking the exterior derivative of \eqref{BH_dstarF}, 
 that local charge conservation ($d\Je = 0$) 
 is preserved.

Thus, 
 by means of this construction of our fields $\Fem, \axi, \axiSource$, 
 we have satisfied conditions (a)--(d) as follows:  
 (a) from the definition of $\Je$, 
 (b) since $\Fem = d \Aem$, 
 (c) from the definition of $\Je$ in combination 
     with $\Fem$ and $\axi$ being disjoint, 
 and
 (d) from either the definition of $\axiSource$ itself, 
  or \eqref{EG_axiSource}.
The last condition,
 (e), 
 is demonstrated below in \ref{ch_Charge}.

The various fields are illustrated
 in figure  \ref{fig_supp_lines_F_axi} which delineates the
 support of each field at $t=10$,
 as well in the 3--dimensional versions
 of figures \ref{fig_3d_RedBit}--\ref{fig_3d_BothBits}.
A cut at time $t=10$ and $z=0$ is shown
 in figure \ref{fig_supp_F_axi_z0}.
It is also possible to represent the forms
 $\Fem, ~\axiSource, ~\axi, ~\axi \wedge \Fem$ as dots,
 lines and surfaces \cite{Gratus-2017arxiv-picto},
 as described in the Appendix
 and shown on figure \ref{fig_form_lines_F_axi}.

% ---------------------------------------------------------------------
\subsection{Visualisation}
\label{ch_Vizualization}

To visualise this electromagnetic construction,
 it is helpful to consider the different support regions
 at a single moment in time $t$.
Following the conventions in figure \ref{fig_supp_lines_F_axi},
 they form 
 the following shapes:
~
\begin{bulletlist}
\item
 The region where the axion {\axiNameSource} $\axiSource\ne0$
  forms a solid torus with a square cross section,
  as shown in dark red
  in figures \ref{fig_3d_RedBit} and \ref{fig_3d_BothBits}.
\item
 The region where the axion {\axiName} $\axi\ne0$
 forms a thick disk,
 which is bounded and includes the region where $\axiSource\ne0$,
 as shown in figures \ref{fig_3d_RedPinkBit}.
\item
 The region where the electromagnetic field $\Fem \ne 0$
  forms a hollow torus with thick surfaces,
  as shown in blue
  on figures \ref{fig_3d_BlueBit} and \ref{fig_3d_BothBits}.
\end{bulletlist}

% ---------------------------------------------------------------------
\subsection{Breaking of global charge conservation: condition (e)}
\label{ch_Charge}

Since the total charge present before the singularity
 is exactly zero,
 any finite amount afterwards
 indicates that global charge conservation has been violated --
 despite the eminently reasonable starting points
 \ref{enum-solution-F} to \ref{enum-solution-Z}
 and criteria
 \ref{enum-criteria-Maxion} to \ref{enum-criteria-Z}.

We now explicitly calculate $\QFut$
 the total charge over any time slice after the singularity.
This is given by
~
\begin{align}
  \QFut 
=
  \int_{\Real^3} \left( d\star \Fem + \axi \wedge \Fem \right)
= 
  \QF + \Qax
,
\label{BH_dstarFaxi}
\end{align}
where $\QF=\int_{\Real^3} d\star \Fem$ and $\Qax=\int_{\Real^3} \axi\wedge \Fem$.
First,
 let us consider the contribution $\QF$. 
 Such a contribution would appear
 as a current in the blue region of Fig. \ref{fig_supp_lines_F_axi},
 and the result could easily be calculated.
However,
 we can instead
 simply observe that for any fixed $t$ time slice,
 wherein $\Fem$ has compact support,
 that
%[
\begin{align}
  \QF
=
  \int_{\Real^3} d\star \Fem
=
  \int_{\textup{Boundary}} \star \Fem
=
  0
.
\end{align}
%]

This now leaves us to calculate the contribution due to axionic currents, 
 $\QFut = \Qax = \int_{\Real^3} \axi \wedge \Fem$. 
The fact that $\axi \wedge \Fem$ is closed 
 (i.e. $d\axi\wedge \Fem = 0$)
 implies local charge conservation. 
However, 
 since $\axi \wedge \Fem$ is not exact 
 (i.e. it cannot be written as the exterior derivative of a two-form)
 the above argument
 using Stokes' theorem leading to $\QF=0$ does not apply here.
In fact,
 for our constructed $\axi$ and $\Fem$,
 we now show that $\Qax\ne 0$.

Now note that the intersection
 of the supports of $\Fem$ and $\axi$ is when
%[
\begin{align*}
  \norm{z}   \le   \tfrac{1}{10}\,t
,~~
  r    \le   \tfrac{5}{10}\, t
,~~
    \left( \tfrac{2}{10} t \right)^2
  \le
    z^2 + \left( r-\tfrac{4}{10} \right)^2
  \le
  \left( \tfrac{3}{10} t \right)^2
.
\end{align*}
%]
This implies $r < \tfrac{3}{10}\,t$, so that
%[
\begin{align}
  \axi
=
  \funz \left(\frac{z}{t}\right)
  ~
  d\left(\frac{z}{t}\right)
.
\label{EG_def_axi_flat}
\end{align}
%]
Thus
%[
\begin{align*}
  \axi \wedge \Fem
&=
  \frac{1}{t}\,\funz\left(\frac{z}{t}\right)\, dz
  \wedge
  \funA'\left(\frac{\rho}{t}\right) \frac{1}{t^2\rho}
\quad\times
\nonumber
\\
&\qquad
  \left[
    t 
    z
    dz \wedge d\theta 
   + 
    t 
    \left( r - \tfrac{4}{10} \right)
    dr \wedge d\theta
   - 
    \rho dt \wedge d\theta
  \right]
\\&=
  \frac{1}{t^3\rho}\,
  \funz\left(\frac{z}{t}\right)\,
  \funA'\left(\frac{\rho}{t}\right)
\quad \times
\nonumber
\\
&\qquad
  \left[ 
    t
    \left( r - \tfrac{4}{10} \right)
    dz \wedge dr \wedge d\theta
   -
    \rho
    dz \wedge dt \wedge d\theta
  \right]
.
\end{align*}
%]

\newcommand\textequal{%
 \rule[.4ex]{3.5pt}{0.35pt}\llap{\rule[.7ex]{3.5pt}{0.35pt}}}
\def\qsquish{\!}  % for easy removal

\noindent
Integrating for some specified time $t>0$
 gives the axion contribution to the total charge $\QFut$:
%[
\def\ptenth#1{\tfrac{#1}{10}}
\begin{align*}
  \QFut \!
&= \qsquish\qsquish
  \int_{\theta \textequal 0}^{2\pi}
  \int_{z \textequal \ptenth{-t}}^{\ptenth{t}}
  \int_{r \textequal \ptenth{t}}^{\ptenth{3t}}
  \qsquish\qsquish
  \frac{(r-\ptenth{4})}{t^2\rho}
  \funz\qsquish\left(\frac{z}{t}\right)
  \qsquish
  \funA'\qsquish\left(\frac{\rho}{t}\right)
  dz 
  \wedge\qsquish dr
  \wedge\qsquish d\theta
\\&=\!
  2\pi
  \int_{z \textequal -\ptenth{1}}^{\ptenth{1}}
  \int_{r \textequal \ptenth{1}}^{\ptenth{3}}
  \frac{(r-\ptenth{4})}{\rho}
  \funz(z) \,
  \funA'(\rho) \,
  dz\wedge dr
\\&=\!
 2\pi
  \int_{z \textequal -\ptenth{1}}^{\ptenth{1}}
  \int_{r \textequal \ptenth{1}}^{\ptenth{3}}
  \frac{(r-\ptenth{4})}{\sqrt{z^2+(r-\ptenth{4})^2}}
  \funz(z)
\quad\times
\nonumber
\\
&\qquad\qquad\qquad\qquad
  \funA'
  \left(
    \sqrt{z^2 + \left( r-\ptenth{4} \right)^2 }
  \right)
  dz\wedge dr
.
\end{align*}
%]
This is
 clearly independent of $t$
 and so is a conserved quantity for $t>0$.
Now we can see that
 it is possible for $\QFut\ne0$;
 by inspection of the
 integrand we have
 $r-\tfrac{4}{10}<0$, $\funz(z)>0$
 and $\funA'(\rho)<0$ hence
 $\QFut\ne0$. 
The non-zero charge originates from both
 the effusion of $\Fem$ and $\axi$ from the singularity,
 and the fact that $\axi$ is not exact.

As a result,
 our claim,
 which would normally be extremely contentious,
 is that total charge is not {globally} conserved,
 i.e. our criterion \ref{enum-criteria-Qe}
 can be met --
 at least in the context of the electromagnetic model
 presented here.
Our conclusion is that electromagnetic models
 do not necessarily enforce charge conservation;
 and this can be said
 not merely as an abstract claim \cite{Gratus-KM-2019foop-nocharge},
 but one grounded in the extant mathematical solution
 we present here.
It remains possible, 
 however, 
 that global charge conservation could be enforced
 by other physical mechanisms.

% ---------------------------------------------------------------------
\subsection{Visualisation in time}
\label{ch_Vizualizationtime}

Although we have specified the mathematical construction
 of our non-conserving solution,
 it is worthwhile describing in more visual terms
 what such a construction would look like.

This is,
 in fact,
 relatively easy.
We have already seen the toroidal structure of the axion {\axiNameSource}es
 and field patterns required
 in figures \ref{fig_3d_RedBit}--\ref{fig_3d_BothBits},
 but only at a single instance in time.
Turning this into a dynamic picture now
 requires us only to observe that
 the arguments of the bump functions
 used in \eqref{EG_def_axi} and \eqref{EG_def_A}
 all contain the factor $r/t$.
Thus any point notionally fixed on the axion {\axiNameSource} distribution
 must hold $r/t$ constant;
 as time passes and $t$ increases,
 $r$ also gets larger.
The toroidal construction therefore expands,
 sized in fixed
 proportion to the spatial cross-section
 of the future causal cone of the singularity.

Although we have constructed our solution
 as if the combined axionic and electromagnetic fields 
 were \emph{emerging} from the singularity,
 it is easily recast in a time-reversed form
 where the construction is instead 
 shrinking and vanishing \emph{into} the singularity.

In appendix \ref{ch_Supp}
 we see how to visualise the fields $Z_S$,
 $\axi$,
 and $\Fem$ 
 as submanifolds.

%=======================================================================
\section{A temporary singularity and transformation electromagnetism}\label{ch_TBH}

% -  -  -  -  -  -

\begin{figure}
\centering
%\resizebox{0.900\columnwidth}{!}{\input{fig13-diffeointro.tex}}
\resizebox{0.900\columnwidth}{!}{\includegraphics{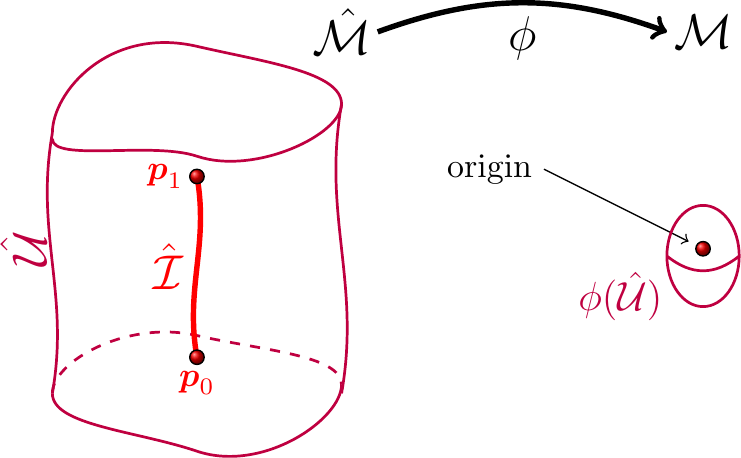}}
\caption{The diffeomorphism $\phi:\Mmanhat\to\Mman$ 
 which maps a topological 3--sphere $\manUHat$
 surrounding the temporary singularity
 to a topological 3--sphere $\phi(\manUHat)$ surrounding the origin.}
\label{fig_map}
\end{figure}

% -  -  -  -  -  -

\begin{figure}
\centering
%\resizebox{0.950\columnwidth}{!}{\input{fig14-diffeocone.tex}}
\resizebox{0.950\columnwidth}{!}{\includegraphics{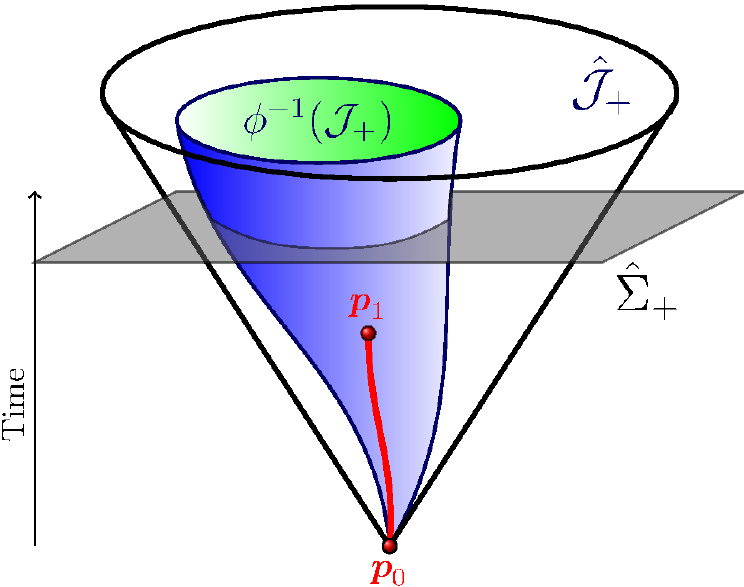}}
\caption{The domains in the morphed spacetime $\Mmanhat$.
The distorted {\colLiCone} cone $\phi^{-1}(\FutLC)$ is
 the causal cone pre-image of the causal cone cone $\FutLC\subset\Mman$. 
This lies inside the future causal cone $\hat{\cal J}_+$
 of the temporary singularity, 
 which is shown as an outlined white cone.
The temporary singularity, 
 which is not part of $\Mmanhat$,
 is indicated by the {\colPsing} curve between $\pini$ and $\pfin$.
Here,
 $\hat\Sigma_+$ is an arbitrary hypersurface intersecting $\FutLChat$
 away from the singularity.
}
\label{fig_hatDom}
\end{figure}

% -  -  -  -  -  -

We would now like to extend the results of the previous sections, 
 which treated a point-like singularity,
 to allow for a singularity that lives for a finite time.

Let $\Mmanhat$ with metric $\ghat$ be a spacetime
 which represents a temporary singularity
 as shown in Fig. \ref{fig_singular}.
We assume 
 that a diffeomorphism $\phi:\Mmanhat\to\Mman$
 can be constructed such that 

\begin{enumerate}[label=(\greek*)]

\item \label{enum-diffeo-item1} % [(\textalpha)]
Any topological 3--sphere
 which surrounds the temporary singularity in $\Mmanhat$
 is mapped to a topological 3--sphere
 which surrounds the instantaneous singularity in $\Mman$; 
 see figure \ref{fig_map}.

\item \label{enum-diffeo-item2} % [(\textbeta)]
The pre-image $\phi^{-1}(\FutLC)$ of the forward causal cone
 of the origin in $\Mman$
 is enclosed within the forward causal cone $\FutLChat$ of 
 the temporary singularity; 
see figure \ref{fig_hatDom}.

\end{enumerate}

% -  -  -  -  -  -

%\def\radcol{cyan}
\def\radcol{gray!50!blue!60}
\def\thetacol{gray!50!red!60}

\begin{figure}
%\resizebox{0.990\columnwidth}{!}{\input{fig15-diffeotime.tex}}
\resizebox{0.990\columnwidth}{!}{\includegraphics{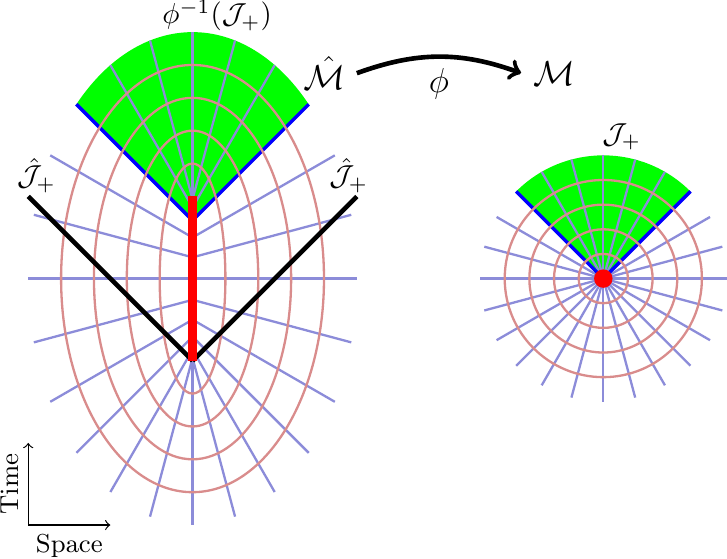}}
\caption{The diffeomorphism $\phi$,
 given by \eqref{TBH_eg_diffeo_timelike}
 which leads to a timelike singularity
 indicated by the {\colPsing} line. 
Here $\FutLChat$ is bounded by the thick black lines
 whereas $\phi^{-1}(\FutLC)$ is given by the {\colInCone} segment. 
The thick black lines are parallel to the boundary of the
  $\phi^{-1}(\FutLC)$.
The blue-gray radial lines in $\Mmanhat$
 are mapped to the radial lines on $\Mman$, 
 likewise the light red ellipses in $\Mmanhat$
 are mapped to circles on $\Mman$.
It is crucial to note that whilst $\phi$
 maps $\Mmanhat$ to $\Mman$, 
 the timelike line singularity is not contained within $\Mmanhat$, 
 nor is the point singularity contained within $\Mman$, 
 and so the mapping remains consistent with $\phi$
 being a diffeomorphism.
}
\label{fig_eg_timelike_sig}
\end{figure}

% -  -  -  -  -  -

An example of such a diffeomorphism
 is given when $\Mman$ is Minkowski spacetime excluding the origin,
 and $\Mmanhat$ is Minkowski spacetime
 excluding the line $\intvHat=\Set{(\tHat,0,0,0),-1\le \tHat\le 1}$. 
The interval $\intvHat$ would be timelike in Minkowski spacetime, 
 and
 the diffeomorphism $\phi:\Mmanhat\to\Mman$ is given implicitly by
%[
\begin{align}
\begin{gathered}
  \tHat
= 
 t
 ~
 \left[ 1 + \left( t^2+x^2+y^2+z^2 \right)^{-1/2} \right]
,
 \quad \xhat=x
,
\\
  \yhat = y
\quadand
  \zhat = z
,
\end{gathered}
\label{TBH_eg_diffeo_timelike}
\end{align}
%]
 as depicted in figure \ref{fig_eg_timelike_sig}.
This diffeomorphism demonstrates
 that even though we cannot use a diffeomorphism
 to map a point into a line,
 we can nevertheless map the region around a point,
 into a region around a line.

\begin{figure}
%\resizebox{0.990\columnwidth}{!}{\input{fig16-diffeospace.tex}}
\resizebox{0.950\columnwidth}{!}{\includegraphics{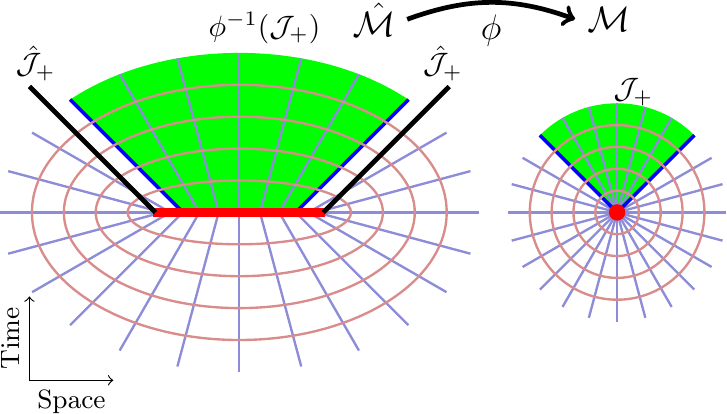}}
\caption{The diffeomorphism $\phi$,
 given by \eqref{TBH_eg_diffeo_spacelike},
 which leads to a spacelike singularity, 
 given by the {\colPsing} line. 
The conventions are as in figure
  \ref{fig_eg_timelike_sig}. }
\label{fig_eg_spacelike_sig}
\end{figure}

An example with a spacelike temporary
  singularity by $\Mmanhat=\Real^4\backslash\intvHat$,
 $\intvHat=\Set{(0,\xhat,0,0),-1\le \xhat\le 1}$ is
  $\phi:\Mmanhat\to\Mman$
 given implicitly by
%[
\begin{align}
\begin{gathered}
\tHat=t
,\quad 
  \xhat
=
  x
  ~
  \left[ 1 + \left( t^2+x^2+y^2+z^2 \right)^{-1/2} \right]
,
\\
  \yhat = y
\quadand
  \zhat = z
,
\end{gathered}
\label{TBH_eg_diffeo_spacelike}
\end{align}
%]
 as depicted in figure \ref{fig_eg_spacelike_sig}.

% -  -  -  -  -  -

% -  -  -  -  -  -

In the general case,
 with $\phi$ satisfying
 \ref{enum-diffeo-item1} and \ref{enum-diffeo-item2} above, % (\textalpha) and (\textbeta)
 let $\starhat$ be the Hodge dual in $\Mmanhat$ corresponding to
 the metric $\ghat$,
 and let $\Fhat\in\Gamma\Lambda^2\Mmanhat$,
 $\axihat\in\Gamma\Lambda^1\Mmanhat$,
 $\axiSourcehat\in\Gamma\Lambda^2\Mmanhat$ and
 $\Jhat\in\Gamma\Lambda^3\Mmanhat$
 be the corresponding fields/sources on
 $\Mmanhat$ given respectively by
%[
\begin{equation}
\begin{gathered}
  \Fhat = \phi^\star \Fem
,\quad
  \axihat = \phi^\star \axi
,\quad
  \axiSourcehat = \phi^\star \axiSource
\\
\qquadand
  \Jhat = \phi^\star \Je + d\left( \starhat \phi^\star \Fem - \phi^\star \star \Fem \right)
\end{gathered}
\label{Mhat_fields}
\end{equation}
%]
where $\phi^*: \Gamma\Lambda^2\Mmanhat \rightarrow \Gamma\Lambda^2\Mmanhat$
 is the pullback of $\phi$,
 which,
 we note,
 commutes with the exterior derivative \cite{BaezMunain-GFKG}. 
The latter two terms on the right hand side of the equation for $\Jhat$
 ensure that $\Fhat$,
 $\axihat$,
 and $\Jhat$ satisfy an equation analogous to \eqref{BH_dstarF}.
In fact,
 all the induced fields satisfy equations analogous to
 \eqref{BH_dstarF}, \eqref{BH_dF},
 \eqref{BH_consv_Charge}, \eqref{BH_axi_Source},
 and \eqref{BH_zetaS_F_abhor}, i.e.
%[
\begin{equation}
\begin{gathered}
  d\starhat \Fhat + \axihat\wedge \Fhat = \Jhat
,\quad
  d\Fhat = 0
,\quad
  d\Jhat=0
,
\\
  d\axihat = \axiSourcehat
\quadand
\axiSourcehat\wedge\Fhat=0\,.
\end{gathered}
\label{TBH_Fields_eqns}
\end{equation}
%]
Let $\SigmaFutHat$ be a spacelike hypersurface
 which intersects $\FutLChat$
 and is away from the temporary singularity.  
Then, 
 from \eqref{BH_current_int_U}
 and the fact that integrals are preserved under diffeomorphism,
 we have
%[
\begin{equation}
\begin{aligned}
  \QFutHat
&=
  \int_{\SigmaFutHat} \Jhat
=
  \int_{\manUHat} \Jhat
=
  \int_{\manUHat}
  \left[
    \phi^\star \Je
   + 
    d \left( \starhat\phi^\star F-\phi^\star\star \Fem \right)
  \right]
\\
&=
  \int_{\manUHat}
  \phi^\star \Je
 +
  \int_{\partial\manUHat} \left( \starhat\phi^\star \Fem - \phi^\star\star \Fem \right)
=
  \int_{\hat{\manU}} \phi^\star \Je
\\
&=
  \int_\manU \Je
=
  \QFut
 \ne
  0
.
\end{aligned}
\label{TBH_Qhat_+}
\end{equation}
%]
where $\hat{\manU}$ is a topological 3--sphere which surrounds the
singularity and $\manU=\phi(\hat{\manU})$.

Again we have satisfied
 \ref{enum-solution-F}--\ref{enum-solution-Z}         % (i)--(iii)
 and   
 \ref{enum-criteria-Maxion}--\ref{enum-criteria-Qe}       % (a)-(e)
 and we conclude that the global charge is
 not conserved in the more general spacetime $\Mmanhat$.

%=======================================================================
\section{Conclusion}\label{ch_CON}

In this article, 
 we have investigated the behaviour of charge conservation, 
 a usually sacrosanct principle 
 of standard electromagnetism.
It has already been shown that 
 it is possible to break global charge conservation
 whilst preserving local charge conservation
 \cite{Gratus-KM-2019foop-nocharge},
 and this was done by considering 
 an extension of Maxwell's equations where
 the excitation fields $\VD$ and $\VH$ are no longer physical fields,
 and placing this in a spacetime
 containing a temporary singularity.
This situation is in common with other consequences of singularities,
 where it is said that since ``physics breaks down'',
 anything might occur as a result. 
However,
 bearing in mind the
 H.G. Wells quote:
 ``If anything is possible, then nothing is interesting'',
 we were motivated to find examples that constrain this
 inchoate sense of the possibilities.

Here we have substantiated a situation
 where a \cRed{minimally extended} Maxwellian electromagnetism with an axionic coupling 
 \cite{Gratus-MK-2020pra-area51}
 in a topologically non-trivial spacetime
 fails to conserve global charge, 
 by
 mathematically specifying the required space and time dependence
 of the relevant fields.
In this construction there are no fields
 before the advent of the singularity,
 and the axions and charge emerge from it; 
 the scheme can even be time reversed so as to destroy 
 correctly configured and collapsing arrangements 
 of axions and charge.
Attempts to recover global charge conservation
 by accounting for the axionic charge ``in'' the singularity
 (before it emerges into the universe)
 are bound to fail as we have deliberately restricted attention
 to situations 
 that are diffeomorphic to an instantaneous singularity.
Further, 
 our specification can be transformed by diffeomorphism
 to apply also to a range of related scenarios, 
 including that for singularity
 that lives for a finite time.
In doing this we have also demonstrated that although
 in a general sense 
 anything might happen at a singularity, 
 in practise \emph{anything} will not.
This is because the spacetime surrounding the singularity
 is subject to physical law, 
 which constrains the means by which the ``anything'' can happen.

It is, 
 however, 
 certainly arguable that 
 the form of the solutions is somewhat contrived,
 i.e. that no such field configuration will form randomly
 or be created naturally.
However, in the spirit of Morris and Thorne's famous paper on wormholes
 \cite{Morris-T-1988ajp-wormholes}
 we can ask
 ``What constraints do the laws of physics place
 on the activities of an arbitrarily advanced civilization?''
Suppose an advanced civilization feels it is necessary
 to adjust the total charge of the universe,
 having predicted -- or perhaps manufactured -- 
 the arrival of a temporary singularity.
They will then be able to construct
 the fields $\Fem$, $\axiTop$ and $\axiSource$
 in such a way that they will all vanish,
 along with axionically driven charge
 {($Q_-= \int_{\SigmaPast}\axi\wedge \Fem$)},
 into singularity.

Our conclusion appears to be at once startling and undeniable: 
 global charge conservation cannot be guaranteed
 in the presence of 
 axionic electromagnetic interaction.

\acknowledgments
   Both JG and PK are grateful for the support provided by
 STFC (Cockcroft Institute ST/G008248/1 and ST/P002056/1)
 and
 EPSRC (Alpha-X EP/N028694/1). 
PK would also like to acknowledge recent support
 from the UK National Quantum Hub for Imaging (QUANTIC, EP/T00097X/1).

%=======================================================================
%\bibliographystyle{unsrt}
%\bibliography{../../../bibtex.bib}
%\bibliography{bh17pk-export}

%=======================================================================
%=======================================================================
%\clearpage
%\newpage
\appendix

\section{Depicting forms as curves and surfaces.}
\label{ch_Supp}

% -  -  -  -  -  -

\begin{table}
\centering
\begin{tabular}{|c|c|c|c|c|}
\hline
Form & Degree & Parity & Submanifold on timeslice & Colour %Depiction
\\\hline
$\Fem$ & 2 & untwisted & 1d circles in $(r,z)$ plane & blue %circles
\\\hline
$\axi$ & 1 & twisted & 2d disc in $(r,\theta)$ plane & light red %discs
\\\hline
$\axiSource$ & 2 & twisted & 1d circles in $(r,\theta)$ plane & dark red %circles
\\\hline
$\Fem \wedge \axi$ & 3 & twisted & 0d dots & green %diamonds
\\\hline
\end{tabular}
\caption{The four fields $\Fem$, $\axi$, $\axiSource$ and $\Fem \wedge \axi$
  and how they are depicted in figures \ref{fig_form_lines_F_axi} and
  \ref{fig_lines_F_axi_z0}.
}
\label{tab_forms_depiction}
\end{table}

% -  -  -  -  -  -

\begin{figure}
%\resizebox{0.900\columnwidth}{!}{\input{fig17-jgthetaslice.tex}}
\resizebox{0.900\columnwidth}{!}{\includegraphics{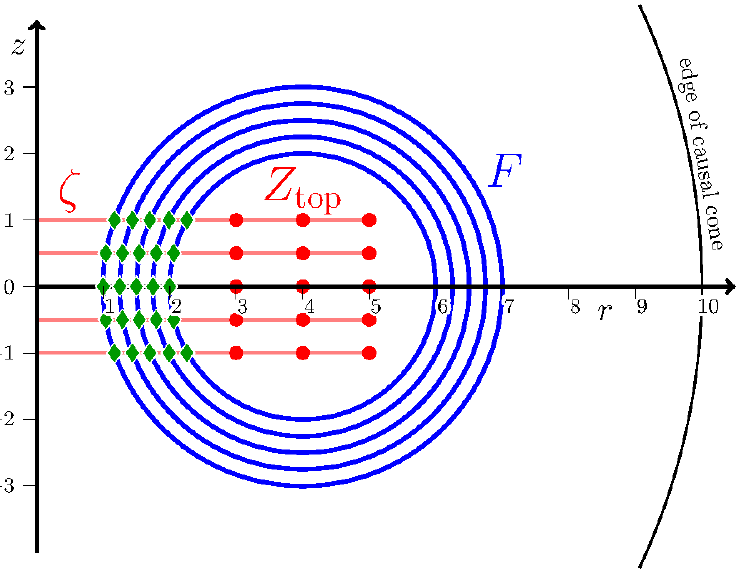}}
\caption{Representing the four fields
 on a slice at $\theta=0$ at $t=10$, 
 showing the $r,z$ plane.
Here the timeslice submanifolds of $\Fem$
 overlap the $\theta=0$ plane, 
 and so are shown as blue circles, 
 as per table \ref{tab_forms_depiction}.
The two axionic fields --
 $\axi$ (light red)
 and
 $\axiSource$ (dark red) --
 have timeslice submanifolds that intersect with the $\theta=0$ plane,
 and so are depicted as lines and dots.
Lastly, 
 the points of the timeslice submanifold for
 $\axi\wedge \Fem$
 are shown as green diamonds.
This is the same view as on figure \ref{fig_supp_lines_F_axi}.
}
\label{fig_form_lines_F_axi}
\end{figure}

% -  -  -  -  -  -

\begin{figure}[t]
\centering
%\resizebox{0.900\columnwidth}{!}{\input{fig18-jgzslice.tex}}
\resizebox{0.900\columnwidth}{!}{\includegraphics{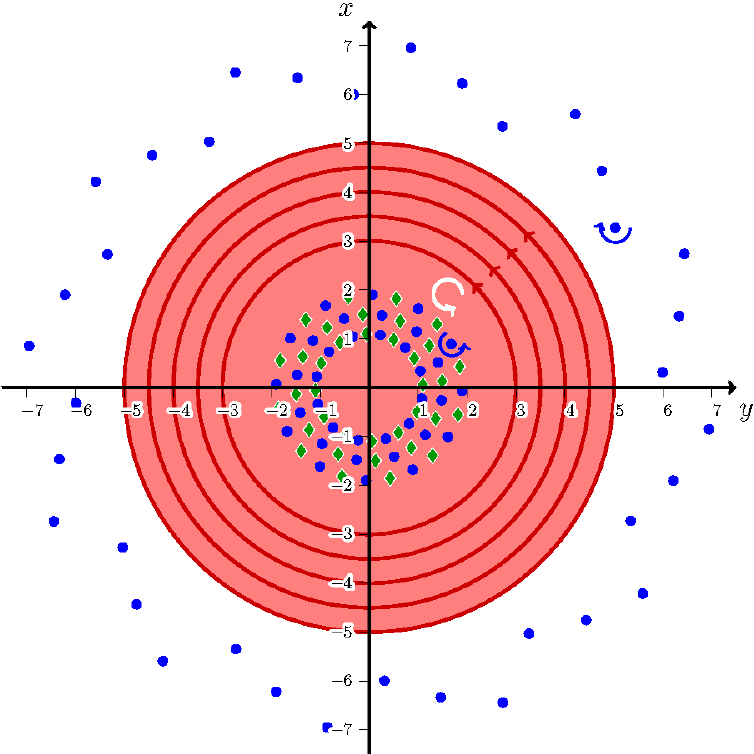}}
\caption{Representing the four fields
 on a slice $z=0$ at $t=10$, 
 showing the $x,y$ (or $r,\theta$) plane.
Here $\Fem$
 has timeslice submanifolds that intersect with the $z=0$ plane, 
 and so are depicted as blue dots.
The two axionic fields --
 $\axi$ (light red)
 and
 $\axiSource$ (dark red) --
 have timeslice submanifolds that overlap the $z=0$ plane, 
 and so are shown as a disc and circles
 as per table \ref{tab_forms_depiction}.
Lastly, 
 the points of the timeslice submanifold for
 $\axi\wedge \Fem$
 are shown as green diamonds.
The orientations of $\Fem$
 on this diagram 
 are indicated by blue arrows curling around selected dots, 
 and the orientation of $\axi$ 
 by a white curled arrow.
This is the same view as on figure \ref{fig_supp_F_axi_z0}.
}
\label{fig_lines_F_axi_z0}
\end{figure}

% -  -  -  -  -  -

It can be extremely useful to
 represent forms pictorially by dots,
 curves and surfaces \cite{Gratus-2017arxiv-picto}, 
 because once the conventions are learnt, 
 it enables information to be conveyed
 more easily and more accurately.
In $4$ dimensions a 1--form is given by 3--dimensional volumes,
 a 2--form by 2--dimensional surfaces
 and a 3--form by 1--dimensional curves. 
Closed forms 
 are submanifolds that do not have boundaries,
 while non-closed forms do have boundaries.

These submanifolds (forms) have an orientation, 
 where
 untwisted forms have an external orientation
 while twisted forms have internal orientation.
To show these submanifolds
 for our field solution, 
 we have used two slices. 
Both slices are at $t=10$,
 but with $\theta=0$ in figure \ref{fig_form_lines_F_axi},
 contrasting with 
 an orthogonal slice at $z=0$ in figure \ref{fig_lines_F_axi_z0}. 
Figure \ref{fig_lines_F_axi_z0}
 also shows the orientations of the fields.
In table \ref{tab_forms_depiction},
 we list the four fields $\Fem$,
 $\axi$,
 $\axiSource$,
 and $\Fem\wedge\axi$.
We show how 
 they are depicted in figures \ref{fig_form_lines_F_axi} and
  \ref{fig_lines_F_axi_z0}.

For the timeslice $t=10$, 
 the electromagnetic 2--form $\Fem$ is a set of closed surfaces, 
 and on the intersection with $\theta=0$,
 are represented by circles, 
 as can be seen on figure \ref{fig_form_lines_F_axi}.
These circles lie in the $(r,z)$ plane
 and on the intersection with $z=0$, 
 $\Fem$ is represented as dots,
 as depicted on figure \ref{fig_lines_F_axi_z0}.
Since $\Fem$ is an untwisted form,
 their orientation is external
 and on figure \ref{fig_lines_F_axi_z0}
 can be shown as an arrow which curls round the blue $\Fem$ dots.
Since the blue $\Fem$ circles on figure \ref{fig_form_lines_F_axi}
 come out of the $z=0$ plane in the region $6<r<7$,
 and go into the plane in the region $1<r<2$,
 they have opposite orientations in these two regions.
These blue circles also mimic the magnetic field lines.
The electric field is then perpendicular to these circles, 
 pointing towards the next circle 
 of the same radius at greater $\theta$.

The 2--form axion {\axiNameSource} $\axiSource$,
 depicted in dark red,
 is represented by circles in the $(r,\theta)=(x,y)$ plane. 
These are twisted,
 so have an internal orientation
 which is shown as anticlockwise in figure \ref{fig_lines_F_axi_z0}.

The 1--form axion {\axiName} $\axi$,
 depicted in light red,
 is represented by discs in the $(r,\theta)$ planes. 
These are twisted,
 so have an internal orientation
 which is shown as shown (in white)
 as anticlockwise in figure \ref{fig_lines_F_axi_z0}.
This is because it is compatible with the orientation
 of $\axiSource$.

The 3--form $\axi\wedge \Fem$, depicted as green diamonds. 
These diamonds are in the region where
 the supports of $\axi$ and $\Fem$ intersect.
Since the orientation of $\axi$ and $\Fem$ are the same in this region,
 the orientation of $\axi \wedge \Fem$ is simply ``$+$''. 
In spacetime this 3--form is depicted as lines, 
 i.e. straight line flowing outward from the origin. 
The internal orientation are arrows pointing out of the page.

\begin{figure}
\centering
\resizebox{0.980\columnwidth}{!}{\includegraphics{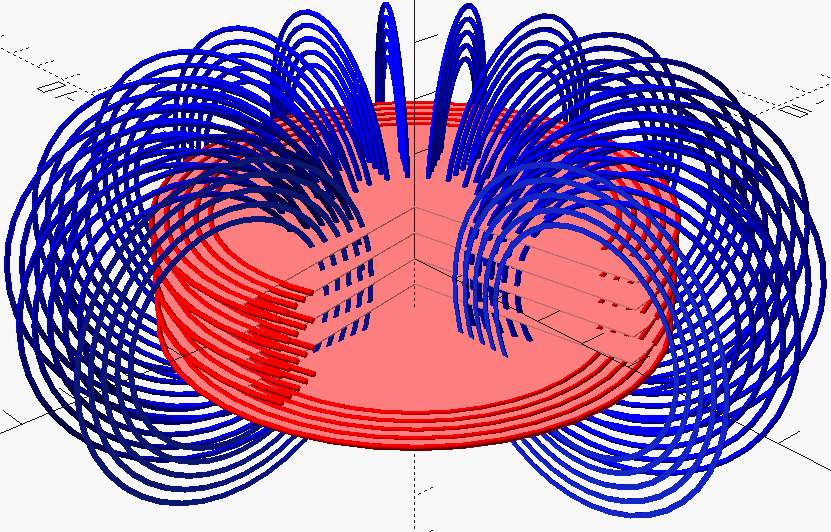}}
\caption{A three dimensional representation of the $\Fem$ (blue), 
 $\axi$ (light red), 
 and $\axiSource$ (dark red)
 submanifolds 
 on a timeslice.
Each of the three field elements has a 90$^\circ$ wedge cut out,
 with each wedge rotationally offset by a small angle, 
 to help show and clarify interior detail.
}
\label{fig_lines_F_axi_3D}
\end{figure}

A three dimensional view of the field forms
 is shown on figure \ref{fig_lines_F_axi_3D}.
Indeed, 
 this visualization is particularly useful when 
 considering how such a field configuration
 could be generated.
We see that in the snapshot, 
 the $\VB$ field lines need to form a cylinder 
 of field loops bent around into a torus.
Such an arrangement might be achieved using electric
 currents passing in opposite directions along 
 a pair of hollow concentric wires,
 with the magnetic field thus localized between them.
Conveniently, 
 the electric field $\VE$ can then arise naturally, 
 because the solution -- 
 and hence the tori --
 are expanding with time, 
 and that time-dependent $\VB$ field directly produces $\VE$.
Further, 
 we can see %at least 
 that the axion flux $\axiSource$ is oriented 
 along loops around the main axis, 
 representing a circulating flux
 consistent with the axion field $\axi$.

% -  -  -  -  -  -

%=======================================================================

\section{Proof about the future causal cone $\FutLC$}
\label{ch_Proofs}

\begin{proof}[Proof of \eqref{BH_Fut_MmanSup}.]
Observe that if $p\in\FutLC^{\text{sup}}(\interval)$ then there is a
causal curve connecting $\interval$ to $p$. This will intersect every
3--sphere $\manU$ between $\interval$ to $p$. By contrast if
$p\not\in\FutLC^{\text{sup}}(\interval)$ then there exists a 3--sphere
surrounding $\interval$ which does not intersect the backward causal
cone of $p$.
\end{proof}

%=======================================================================

\section{Remarks on a Lagrangian}
\label{ch_Lagrangian}

Another interesting question is 
 whether or not it is easy to construct a Lagrangian for our topological axion field: 
 the presence of a Lagrangian would assure us that our 
 model has features convenient for a wider context, 
 such as 
 the derivation of the corresponding stress-energy tensor, 
 consequent conservation laws, 
 and even path-integrals.
A Lagragian
 can be constructed if we
 also assume that $\Aem \wedge \axiSource = 0$,
 where $\Aem$ is the electromagnetic potential\footnote{The constraint
  $\Aem \wedge \axiSource=0$ is, 
  however, 
  harder to \emph{derive} from an action.}.
With this  Lagragian as the integrand,
  \eqref{BH_dstarF} follows by varying the action
  %given by
  %[
  \begin{align}
  S[\Aem] = \int \left(\tfrac12 d\Aem \wedge \star d\Aem - \Aem\wedge \Je +
  \tfrac12 \Aem \wedge \axi \wedge d\Aem \right)
\label{Appx_Lagragian}
  \end{align}
  %]
  with respect to $\Aem$,
 where $\Fem=d\Aem$.

%\begin{proof}
To motivate this Lagrangian, 
 note that when 
 varying $S$ with respect to $A$ we have
%[
\begin{align*}
\delta S
&=
 \int \left(
\tfrac12 d\delta\Aem \wedge \star d\Aem 
+
\tfrac12 d\Aem \wedge \star d\delta\Aem  \right.
\\&\qquad\qquad \left.
-\delta\Aem\wedge \Je 
+
\tfrac12 \delta\Aem\wedge \axi \wedge d\Aem 
+
\tfrac12 \Aem\wedge \axi \wedge d\delta\Aem 
\right)
\\&=
 \int \left(
  d\delta\Aem \wedge \star d\Aem 
-\delta\Aem\wedge \Je 
+
\tfrac12 \delta\Aem\wedge \axi \wedge d\Aem 
\right.
\nonumber
\\
&\qquad\qquad
  \left.
  -
   \tfrac12 d(\Aem\wedge \axi) \wedge \delta\Aem 
  \right)
\\&=
 \int \left(
\delta\Aem \wedge d\star d\Aem 
-\delta\Aem\wedge \Je 
+
\tfrac12 \delta\Aem\wedge \axi \wedge d\Aem 
\right.
\nonumber
\\
&\qquad\qquad
  \left.
-
\tfrac12 d\Aem\wedge \axi \wedge \delta\Aem 
-
\tfrac12 \Aem\wedge d\axi \wedge \delta\Aem 
\right)
\\&=
  \int 
    \delta\Aem \wedge 
  \left( 
    d\star d\Fem 
   -
    \Je 
   +
    \axi \wedge d\Aem 
   +
    \tfrac12 \Aem\wedge \axiSource 
  \right)
.
\end{align*}
%]
However, as stated, 
 although we require that $\Aem\wedge\axiSource=0$, 
 it is impotant to note that,
 since $\Aem$ is a potential,
 it is not globally defined. 
Thus we can only assume we can find a topologically trivial region
 of $\Mman$ where $\Aem$ is defined such that  $\Aem\wedge\axiSource=0$. 
In this case \eqref{BH_dstarF}, 
 our axionic Maxwell-Amp{\`e}re-Gauss equation 
 \emph{does} follow from \eqref{Appx_Lagragian}.
Although we do not claim that it is always possible to find such a region, 
 in the solution presented here 
 $\Fem\wedge\axiSource=0$,
 and we can therefore locally choose gauges where $\Aem\wedge\axiSource=0$.
%\end{proof}

%=======================================================================

\section{Spacetime metric and the Proof of \eqref{BH_zetaS_F_abhor_31}}
\label{ch_MetricProof}

\begin{proof}
Since
%[
\begin{align*}
&  \axiSE\wedge\dual{V}\wedge \star \left(  B\wedge \dual{V} \right)
\nonumber
\\
&\quad=
  \axiSE\wedge\dual{V}\wedge i_V \star B
=
  i_V (\axiSE\wedge\dual{V}) \wedge \star B
\\
&\quad=
 -
  \axiSE\wedge (i_V \dual{V})\wedge \star B
=
  \axiSE \wedge \star B
=
  g(\axiSE,B)\star 1
.
\end{align*}
%]
Likewise
%[
\begin{align*}
  & \star \left( dt \wedge \axiSB \right) \wedge
  (dt \wedge E)
\\
&\quad=
  (dt \wedge E)\wedge \star \left( dt \wedge \axiSB \right)
=
  g(\axiSB,E) \star 1
.
\end{align*}
%]
Also from the star pivot
%[
\begin{align*}
  \star \left( dt \wedge \axiSB \right) 
  \wedge
  \star \left( dt \wedge \axiSB \right)
&=
  \left( 
    dt \wedge
    \axiSB 
  \right)
  \wedge \star \star 
  \left( dt \wedge \axiSB \right) 
\\
&=
 -
  \left( dt \wedge \axiSB \right)
  \wedge
  \left( dt \wedge \axiSB \right) 
=
  0
.
\end{align*}
%]
From \eqref{BH_zetaS_F_abhor} we have
%[
\begin{align*}
  0
&=
  \axiSource\wedge \Fem
\nonumber
\\
&=
  \left(
    dt \wedge \axiSE
   +
    \star \left( dt \wedge \axiSB \right)
  \right)
  \wedge
  \left(
    dt \wedge E
   +
    \star \left( dt \wedge B \right)
  \right)
\\
&=
  dt \wedge \axiSE \wedge \star 
  \left( dt \wedge B \right)
 +
  \star
  \left( dt \wedge \axiSB \right)
  \wedge
  \left(dt \wedge E \right)
\nonumber
\\
&\qquad
 +
  \star 
  \left( dt \wedge \axiSB \right) 
  \wedge \star
  \left( dt \wedge \axiSB \right)
\\
&=
  \left(
    g(\axiSB,E) + g(\axiSE,B)
  \right)
   \star 1
.
\end{align*}
%]
%
%
\end{proof}

\clearpage

\widetext

\section{Popular summary}\label{S-popular}

\large

\begin{center}
{\huge{What is an Axion Bomb?}}\\
~\\
\emph{\small{A light-hearted popular summary for ``Temporary Singularities and Axions: \\
 an analytic solution that challenges charge conservation''}}\\
~\\
{Paul Kinsler}
\end{center}

\setlength{\parskip}{1ex}

``Physics breaks down at a singularity'' 
 is one of the most famous statements in pop-physics.
By showing how this might be done\footnote{See
     {{\lq\lq}Temporary Singularities and Axions: \\
 an analytic solution that challenges charge conservation{\rq\rq}},\\
      J. Gratus, P. Kinsler, M.W. McCall, \qquad
        {Annalen der Physik \textbf{533}, 2000565 (2021)},\\
      {http://doi.org/10.1002/andp.202000565}, \qquad
        {https://arxiv.org/abs/2008.02135}.}, 
   one of the most cherished laws of physics, the conservation of charge,
   has come under fire in a recent paper by three physicists:
   Jonathan Gratus from Lancaster,
   and 
  Paul Kinsler and Martin McCall, from Imperial.
``By dropping an `axion-bomb' into a 
   temporary singularity, such as an evaporating black hole, 
  we can create or destroy electrical charge'', claims Kinsler.
Axions are a  hypothesised particle that are a candidate for dark matter, 
 although their exact properties are still debated, 
 and they have not yet been detected.

McCall explains further -- ``This so-called axion-bomb, 
 as just referred to by my somewhat over-enthusiastic colleague,  
 is a mathematical construct that combines electromagnetic fields
 and axion particle fields in the correct way''.
``The construction shinks and disappears into the singularity'', 
 says Gratus,
 ``taking electrical charge with it.
 And it is the combination of a temporary singularity
 and a newly proposed type of axion field that is crucial to its success''.

``There are also philosophical implications'', 
 says Kinsler, 
 ``although people often like to say that `physics breaks down',
 here we show that although exotic phenomena might occur, 
 what actually happens still is very constrained by the 
 still-working laws of physics around the singularity''.
In parting, 
 Gratus interjects ``You know, 
 I first suggested the reverse of this process", 
 he says wistfully, 
 ``and an `axion-fountain' has a much nicer ring to it, 
 don't you think?''.

Paul Kinsler is now safely working on an entirely different project.

\end{document}